\documentclass[11pt,twoside]{article}
\usepackage{atmp}
\usepackage{orbisty}
\begin{document}
\title{Consistency 
of Orbifold Conformal Field Theories on $K3$}
\url{hep-th/0010281}		
\author{Katrin Wendland\footnote{supported by U.S. DOE grant DE-FG05-85ER40219,
TASK A}}
\address{Department of Physics and Astronomy,\\
University of North Carolina at Chapel Hill}
\addressemail{wendland@physics.unc.edu}	
\markboth{\it Consistency of orbifold conformal field theories on 
$K3$}{\it Katrin Wendland}
\begin{abstract}\noindent
We explicitly determine the locations of $G$ orbifold conformal field theories,
$G=\Z_M,\,M\in\{2,3,4,6\}$, $G=\wh D_n,\,n\in\{4,5\}$, or $G$
the binary tetrahedral group $\widehat\T$,
within the moduli space $\m M^{K3}$
of $N=(4,4)$ superconformal field theories associated
to $K3$. This is achieved purely from the known description 
of the moduli space \cite{asmo94}
and the requirement of a consistent embedding of orbifold
conformal field theories within $\m M^{K3}$. 
We calculate the Kummer type lattices
for all these orbifold limits. Our method allows an elementary derivation of
the B--field values in direction of the exceptional divisors that arise
from the orbifold procedure \cite{as95,do97,blin97}, without recourse to
D--geometry. We show that our consistency requirement fixes these values
uniquely and determine them explicitly. The
relation of our results 
to the classical McKay  correspondence is discussed.
\end{abstract}
\section{Introduction}
In this paper, we study certain subvarieties of the moduli space $\cal M$
of $N=(4,4)$ superconformal field theories with central charge $c=6$.
More precisely, all theories in $\cal M$ are assumed to be representations
of the $N=(4,4)$ linear extension of the $N=(2,2)$ superconformal algebra
that contains   $su(2)_l\oplus su(2)_r$, a special
case of the Ademollo et al.\ algebra \cite{aetal76}. Moreover,
with respect to 
a Cartan subalgebra of $su(2)_l\oplus su(2)_r$, left and right
charges (i.e.\ doubled spins) 
of each state in our superconformal field theories
are assumed to
be integral.
The structure of $\cal M$ has already been described in detail
\cite{asmo94,as96,rawa98,di99,nawe00}. Let us summarize its
most important features. 

$\cal M$ decomposes into two   components, 
$\cal M=\cal M^{tori}\cup\cal M^{K3}$. Every theory in $\cal M$ can be assigned
to either the torus or the $K3$ component of the moduli space by
its elliptic genus, which vanishes in the torus case and reproduces the 
geometric elliptic genus of $K3$ otherwise \cite{eoty89,nawe00}. 
Each irreducible component of $\cal M$
is locally described by a Grassmannian 
$\cal T^{4,4+\delta}$ \cite{na86,se88,ce91} 
(see Appendix \ref{grassmann} for notations and 
properties of Grassmannians). 
Here, $\delta=0$ for the torus component and
$\delta=16$ for $K3$. Hence the defining data of
a superconformal field theory in $\cal M$ have been 
encoded by a positive definite
four-plane $x\subset\R^{4,4+\delta}$.
Provisionally, let $X$ denote a complex two-torus or a $K3$ surface, 
depending on which component of the moduli space $x$ belongs to. Then
$\R^{4,4+\delta}\cong H^{even}(X,\R)$, where on cohomology
we always use the scalar product which is
induced by the intersection pairing on $X$.
The four-plane $x$ is now interpreted as subspace of $H^{even}(X,\R)$.
By Poincar\'e duality, $H^{even}(X,\Z)$
is an even self-dual lattice of signature $(4,4+\delta)$ (see Appendix
\ref{lattices} for some mathematical background on lattices). 
Hence by Theorem \ref{onelatt},
$H^{even}(X,\Z)$ is uniquely determined up to lattice automorphisms, 
and we assume that an embedding 
$H^{even}(X,\Z)\hookrightarrow H^{even}(X,\R)$ has been chosen.
Then the four-plane $x\subset H^{even}(X,\R)$ 
is specified by its relative position with respect to 
$H^{even}(X,\Z)$. 

Each theory in $\cal M^{tori}$ has a description as nonlinear sigma model
with target space a complex two-torus. The moduli space of toroidal conformal
field theories had originally been given by Narain \cite{na86} in terms
of the odd torus cohomology. To arrive at the above description
in terms of the even torus cohomology one has to use $SO(4,4)$ triality, 
see \cite{nawe00}. 

For $\cal M^{K3}$ one uses the  isomorphism \req{asmois} with
primitive null vectors $\upsilon,\upsilon^0\in H^{even}(X,\Z)$,
$\skp{\upsilon}{\upsilon^0}=1$, to show that its parameter space agrees with
the parameter space of nonlinear sigma models with $K3$ target 
\cite{asmo94}. 
Here, $\upsilon,\upsilon^0$ are interpreted as
generators of $H^4(X,\Z)$ and $H^0(X,\Z)$, respectively. This description
equally holds in the torus case \cite{nawe00}. The image
$(\Sigma,V,B)$ of a given four-plane $x$ under \req{asmois} is called
a \textit{geometric interpretation}. Here, the three-plane 
$\Sigma\subset H^2(X,\R)\cong\R^{3,3+\delta}$ 
is interpreted as the subspace of   self-dual two--forms and thus
encodes an Einstein metric
of volume $1$
on $X$. The three-plane $\Sigma$
is specified by its relative position with respect to the
even self-dual lattice $H^2(X,\Z)\subset H^2(X,\R)$
(see Theorem \ref{onelatt}). The parameter $V$
is interpreted as volume of $X$, and $B\in H^2(X,\R)$ denotes the B-field.
We remark that in contrast to higher dimensional Calabi Yau manifolds we 
need not perform a large volume limit in order to study nonlinear
sigma models on $K3$, since here the metric on the moduli space 
does not receive instanton corrections \cite{nasu95}.

Globally, the irreducible components of $\cal M$ are obtained by modding
out a discrete symmetry group from their local descriptions. Namely,
\begin{equation}\label{modsp}
\cal M^\delta
= O^+(H^{even}(X,\Z))\backslash O^+(H^{even}(X,\R))/ SO(4)\times O(4+\delta)
\end{equation}
\cite{na86,asmo94}, up to a subtlety that results in the choice of 
$O^+(H^{even}(X,\Z))$ instead of $O(H^{even}(X,\Z))$ above \cite{nawe00}.
In order to generate the group $O^+(H^{even}(X,\Z))$ one firstly needs  
the \textit{classical 
symmetries} which identify equivalent Einstein metrics thus fixing
$\upsilon,\upsilon^0$. Secondly, we have B-field shifts by
$\lambda\in H^2(X,\Z)$ which induce 
$(\upsilon,\upsilon^0)\mapsto (\upsilon, 
\upsilon^0+\lambda-{\lambda^2\over2}\upsilon)$ and 
for $w$ with $\langle w,\upsilon\rangle=0$ induce
$w\mapsto w-\langle\lambda,w\rangle\upsilon$. 
Thirdly, one can use 
mirror symmetry \cite{asmo94,asmo}, or the Fourier--Mukai transform
$\upsilon\leftrightarrow\upsilon^0$ \cite{nawe00}, 
where the latter approach
appears to be the simpler one.

Given the above description of $\mathcal M$, we 
formulate the aim of this paper as follows:
Consider a superconformal field theory 
in $\cal M^{tori}$, specified
by a four-plane $x_T\in\cal T^{4,4}$, that admits a discrete symmetry 
$G$ which preserves supersymmetry, so $G\subset SU(2)$. 
Then for nontrivial non-translational $G$ the resulting $G$
orbifold conformal field theory is known to belong to $\cal M^{K3}$
(see, e.g., \cite{eoty89} to check the elliptic genera
in the case of cyclic groups $G=\Z_M$, $M\in\{2,3,4,6\}$).
For all possible such actions that do not contain non-trivial 
translations  (by \cite{fu88} this means for
$G=\Z_M$, $M\in\{2,3,4,6\}$,  
$G=\wh D_n$, $n\in\{4,5\}$, and $G=\widehat\T$),   we  
specify the location of the resulting four-plane
$x\in\cal T^{4,20}$ in a way that allows to explicitly read off  a geometric
interpretation of $x$ on the corresponding
$G$ orbifold limit of $K3$. In particular,
we show that a consistent embedding of the subvarieties which contain  
such orbifold 
conformal field theories in $\cal M^{K3}$ fixes the B-field values of the
above geometric interpretation in direction of the exceptional divisors
of the blow up of the orbifold singularities. We determine these
B-field values  explicitly.

As a first step,  in Sect.\ \ref{geometry},
we describe the underlying geometric picture. In other words, 
we specify the locations  of orbifold limits within the moduli space
$O^+(H^2(X,\Z))\backslash\cal T^{3,19}$ of   Einstein metrics  
with volume $1$ on $K3$.
In contrast to the non-compact minimal resolution of $\C^2/G$, 
the components of the
exceptional divisors on $X$ do not generate a primitive sublattice of
$H_2(X,\Z)$. For $G=\Z_2$, 
in \cite{pss71,ni75} it was shown  that they are rather
contained in a finer lattice, the \textit{Kummer lattice}.
We explicitly calculate the generalizations of the
Kummer lattice  to 
all $G$ listed above.
In Sect.\ \ref{quantum} we 
use Theorems \ref{lattemb} and \ref{grass} to
lift these geometric results to the 
``quantum level''. This in particular leads to a short derivation of 
the correct B--field
values for orbifold conformal field theories
which is essentially independent of the technical discussion in 
Sect.\ \ref{geometry}. We are in agreement with previous results by 
Aspinwall \cite{as95},
Douglas \cite{do97} and Blum/Intriligator \cite{blin97}.
We conclude with a summary and discussion of possible further implications
our techniques might have. We briefly point out their 
surprisingly simple relation to the classical McKay 
correspondence.
There are two appendices to present the necessary mathematical
 background on lattices and Grassmannians.
\begin{acknowledgements}
The results presented in this note were initiated in 
joint work with Werner Nahm 
\cite{nawe00}, whom I am also indebted to for numerous fruitful discussions
during his supervisorship of my doctoral studies, which this work is based on.
I like to thank Paul Aspinwall, Claus Hertling, Gerald H\o hn, 
David R. Morrison, Werner Nahm, and Andreas Recknagel for  
helpful comments on earlier versions of this work.
\end{acknowledgements}
\section{Kummer type constructions of $K3$}\label{geometry}
Apart from notations which are introduced
at the beginning of the present section, further results of this work 
can be understood
without the technical details discussed below. In particular, the   
proofs in 
Sect.\ \ref{quantum} are mostly
independent of Prop.\ \ref{kummerlattice}.

Let $T$ denote a complex two--torus with Einstein metric of volume $1$
specified by the positive definite three--plane $\Sigma\subset H^2(T,\R)$
of self dual two--forms.
Assume that $T$ possesses a nontrivial
discrete symmetry $G\subset SU(2)$, such that
the induced action on $\Sigma$ is trivial.  
The variety $T/G$ has a set $\m S$ of singularities 
of ADE type \cite{du34,ar66}, which we assume to be nonempty. 
By $\wt{\m S}\subset T$ we denote the pre-image of $\m S$ in $T$.
The minimal resolution 
$p:X\longrightarrow{T/G}$
produces a $K3$ surface $X$ \textit{in the orbifold limit}. This means that on
$X$ we use the metric which is induced by the flat torus metric and assigns
volume zero to all components of exceptional divisors $p^{-1}(s), s\in\m S$.

Here, we restrict considerations
to groups $G$ that do not contain non-trivial translations.
In \cite{fu88}, such $G$ actions  
have been classified, and our discussion below covers all these cases
(see \req{notations}). 
In fact, this means that we omit only two further 
orbifold constructions of $K3$
where $G$ does contain  translations \cite{primo}, 
\cite[Table 18]{bdhkmms01}.
To locate all such
$G$ orbifold limits within the moduli space of volume $1$ Einstein metrics 
on $K3$, we will determine the
appropriate embedding of the  image of 
$G$ invariant torus forms   $H^2(T,\Z)^G$
in  $H^2(X,\Z)$. Since the geometric data of Einstein metrics on $T$, $X$
are given by 
three--planes in $H^2(T,\R)$, $H^2(X,\R)$ which
are specified by their relative position with respect to these lattices,
such an embedding indeed is all we need to find.

As to notations, for $s\in\m S$ the irreducible components
of the exceptional divisor $p^{-1}(s)$ are rational spheres with
intersection matrix the negative of the Cartan matrix corresponding
to the type of singularity $s$. Their Poincar\'e duals are lattice vectors
in $H^2(X,\Z)$, which in 
this section we denote\footnote{\label{abuse}This is a slight abuse of notation
as we will see in Sect.\ \ref{quantum}, where the $E_s^{(j)}$ are replaced
by $\wh E_s^{(j)}$.}  
$E_s^{(j)}, (E_s^{(j)})^2=-2$, and which span an ADE type root lattice
$\Gamma_s$. We set $\m E_{|G|}:=\bigcup\limits_{s\in\m S}\{ E_s^{(j)} \}$ and
$\Gamma_{|G|}:=\bigoplus\limits_{s\in\m S}\Gamma_s\subset H^2(X,\Z)$.
Note that $\Gamma_{|G|}$ is a root lattice with fundamental system
$\m E_{|G|}$.

On $T$, we choose complex coordinates $z_1,z_2$  compatible with the
Einstein metric and split $T$ into $T=T^2\times \wt T^2$ with elliptic curves
$T^2$, $\wt T^2$. Both curves
are assumed to be $\Z_M$ symmetric, 
but the metric need not
be diagonal with respect to $z_1,z_2$. 
If $M\in\{2,3,4,6\}$,
we consider the algebraic $G=\Z_M$ action, where $\Z_M$ 
is realized as group of $M^{th}$ roots of unity in $\C$:
$$
(z_1,z_2)\in\C^2,\;
\Z_M\ni\zeta:\quad\quad \zeta.(z_1,z_2)=(\zeta z_1,\qu\zeta z_2);
$$
if $T^2= \wt T^2$ and $M\in\{4,6\}$ we also have 
algebraic $\wh D_{M/2+2}$ actions
with additional generator $I$,
$$
(z_1,z_2)\in\C^2:\quad
I.(z_1,z_2)=(-z_2,z_1).
$$
We have to rewrite the results of \cite[Table 9]{fu88} to notice that
the above $\widehat D_4$ action is algebraic as well on   tori
\begin{equation}\label{tetratorus}\begin{array}{rcl} \displaystyle
T_\T \>:=\> \C^2/\Lambda_{D_4}, \e
\Lambda_{D_4}\>:=\>V\spann_\Z 
\left\{ (1,0),\, (i,0),\, (\inv[i+1]{2},\inv[i+1]{2}), \,
(\inv[i+1]{2},\inv[i-1]{2}) \right\},\quad
V\in\R,
\end{array}
\end{equation}
but with different fixed point sets (see \req{notations}).
To distinguish the two cases, the latter   is denoted 
$\widehat D_4^\prime$. On $T_\T$ as in \req{tetratorus} there also
is an algebraic action of the binary tetrahedral group $\widehat\T$ which
is obtained from the $\widehat D_4^\prime$ action with additional generator
$$
(z_1,z_2)\in\C^2:\quad
J.(z_1,z_2)=\inv[i+1]{2}\left(\,i(z_1-z_2)\,,-(z_1+z_2)\,\right).
$$
From the torus geometry it is natural to label $\Z_2$ and $\Z_4$ type fixed
points by vectors $i\in\F_2^4$, whereas $\Z_3$ type fixed points
carry labels  $t\in\F_3^2$ ($\F_p$, $p$ prime, denotes
the unique finite field with $p$ elements). The integral two--forms
dual to the components of
exceptional divisors in the blow up of an ADE singularity are
labeled as follows:
\begin{eqnarray*}
\mathbf{A_1}:\,\Aa,\;\;\mathbf{A_2}:\,\Ab\!\!\!,\;\;\mathbf{A_3}:\,\Ac\!\!\!,
\;\; \mathbf{A_5}:\,\Aee\!\!\!,\e 
\mathbf{D_4}:\,\Da\!\!\!,\;\;\mathbf{D_5}:\,\Db\!\!\!,\;\;\mathbf{E_6}:\,\Ea\e
\!\!\!\!.
\end{eqnarray*}
Recall that a $\Z_m$ type fixed point gives an $A_{m-1}$ type singularity on 
$T/G$, 
whereas $\widehat D_n$ type and $\widehat\T$ type fixed points correspond
to $D_n$ and $E_6$ type singularities, respectively.
Then for the various orbifold limits of $K3$ we have
\begin{equation}\label{notations}
\begin{array}{c||c|l}
G&\Gamma_{|G|}(-1)&\quad\quad \m E_{|G|}\\[3pt]
\hline\hline&&\\[-7pt]
\Z_2&A_1^{16}&E_i, \;\; i\in\F_2^4,\\[5pt]\hline
\Z_3&A_2^{9}&E_t^{(l)}, t\in\F_3^2,\, l\in\{1,2\},\\[5pt]\hline
\Z_4&A_3^4\oplus A_1^6&E_i^{(l)}, i\in I^{(4)}:=\left\{(j,k),\,j,k\in 
\{(0,0),(1,1)\}\right\},\\
&&\quad\quad\quad\quad\quad l\in\{1,2,3\},\\	
&&E_i,\;\;  
i\in I^{(2)}:=\left\{(j,1,0),(1,0,j),\,j\in \F_2^2\right\},\\ 
&&\quad\quad\quad\quad\quad(1,0,1,0)\sim(0,1,0,1),\\[5pt]\hline
\Z_6&A_5\oplus A_2^4\oplus A_1^5&E_0^{(l)},\, l\in\{1,\dots,5\},\\
&&E_t^{(l)}, t\in \left\{(1,0),(0,1),(1,1),(1,2)\right\},\,l\in\{1,2\},\\
&&E_i,\;\;  i\in \left\{(1,0,0,0),(0,0,0,1),\right.\\
&&\quad\quad\quad\quad\quad    \left.(1,0,1,0),(1,0,0,1),
(0,1,0,1)\right\}\\[5pt]\hline
\wh D_4&D_4^2\oplus A_3^3\oplus A_1^2&E_i^{(l)},i\in\left\{0,1\right\},
l\in\left\{1,\dots,4\right\},\\
&&E_i^{(l)}, i\in \left\{(1,1,0,0),(1,0,1,0),(1,0,0,1)\right\},\\
&&\quad\quad\quad\quad\quad l\in\{1,2,3\},\\
&&E_i,\;\;
i\in \left\{(1,0,0,0),(0,1,1,1)\right\},\\[5pt]\hline
\wh D_4^\prime&D_4^4\oplus A_1^3&
E_i^{(l)},i\in I^{(4)}:=\left\{(j,k),j,k\in 
\{(0,0),(1,1)\}\right\}, \\
&&\quad\quad\quad\quad\quad l\in\{1,\dots,4\},\\[2pt]
&&E_i, \;\;
i\in\{(1,0,0,0), (0,0,0,1), (1,0,1,0) \}, \\[5pt]\hline
\end{array}
\end{equation}
\addtocounter{equation}{-1}
\begin{equation}\label{notationstoo}
\begin{array}{c||c|l}
G&\Gamma_{|G|}(-1)&\quad\quad \m E_{|G|}\\[3pt]
\hline\hline&&\\[-7pt]
\wh D_5&D_5\oplus A_3^3\oplus A_2^2\oplus A_1&E_0^{(l)}, l\in\{1,\dots,5\},\\
&&E_i^{(l)}, i\in \left\{(1,0,1,0),(1,0,0,1),(0,1,0,1)\right\},\\
&&\quad\quad\quad\quad\quad l\in\{1,2,3\},\\
&&E_t^{(l)}, t\in \left\{(1,0),(1,1)\right\},\,l\in\{1,2\},\\
&&E_i,\;\; i=(0,0,0,1).\\[5pt]\hline
\wh\T & E_6\oplus D_4\oplus A_2^4\oplus A_1&
E_0^{(l)}, \quad\quad\quad l\in\{1,\dots,6\},\\
&&E_{(1,1,0,0)}^{(l)}, \quad l\in\{1,\dots,4\},\\
&&E_t^{(l)}, \;  t\in \left\{(1,0),(0,1),(1,1),(1,2)\right\},\, l\in\{1,2\},\\
&&E_{(1,0,0,0)}.\\[5pt]\hline
\end{array}
\end{equation}
From the very definition of the orbifold construction we have
a rational map $\pi:T\longrightarrow X$ of degree $|G|$, which is 
defined outside the fixed points of $G$.
By $K_{|G|}$, $\Pi_{|G|}$ we denote the primitive sublattices of
$H^2(X,\Z)$ that contain $\pi_\ast(H^2(T,\Z)^G)$, $\Gamma_{|G|}$,
respectively. From \req{notations} one checks  that 
$\rk K_{|G|}+ \rk \Pi_{|G|}=22=\rk H^2(X,\Z)$, confirming that  $X$
indeed
is a $K3$ surface. $K_{|G|}\perp\Pi_{|G|}$ by construction, since all 
exceptional divisors have volume zero with respect to any Hermitean 
metric compatible with the Einstein metric in the orbifold limit. Hence  by
Theorem \ref{lattemb} there is
an isomorphism $\gamma: (K_{|G|})^\ast/K_{|G|} \rightarrow 
(\Pi_{|G|})^\ast/\Pi_{|G|}$. Moreover, the embedding 
$\pi_\ast(H^2(T,\Z)^G)\hookrightarrow H^2(X,\Z)$ is determined, once
we know the lattices $ K_{|G|}$, $\Pi_{|G|}$, and $\gamma$. The  
rest of this section therefore is devoted to a geometrically motivated 
construction of these data.

Note that $\pi_\ast(H^2(T,\Z)^G)\cong 
H^2(T,\Z)^G(|G|)$ by \cite[Prop.\ 1.1]{in76}. Since we prefer to work with
metric isomorphisms, we denote the $\pi_\ast$ image of $\kappa\in H^2(T,\Z)^G$
by $\sqrt{|G|} \kappa$. This is also in accord with the
fact that $\pi_\ast \pi^\ast=|G|$ and $\pi^\ast\pi_\ast =|G|$  by
\cite[Prop.\ 1.1]{in76}.

Let us recall Nikulin's solution to our problem in the case
$G=\Z_2$, i.e.\ for classical Kummer surfaces. In \cite{pss71,ni75}, 
it is proven
that $K_2\cong H^2(T,\Z)(2)$, and $\Pi:=\Pi_2$, the \textit{Kummer lattice},
is determined to
\begin{equation}\label{kummer}
\Pi = \spann_\Z\{ E_i,\, i\in\F_2^4; \quad
\inv{2} \sum_{i\in H} E_i,\, H\subset \F_2^4 \mbox{ a hyperplane} \}.
\end{equation}
We  interpret the description of $H^2(X,\Z)$
that arises from Theorem 
\ref{lattemb} as follows: 
Clearly, $H^2(X,\Z)$ contains $\Pi$ and  $K_2$.
The latter consists of the Poincar\'e duals $\sqrt2\kappa$ of 
images of torus two--cycles that correspond to $\kappa\in H^2(T,\Z)$. 
These cycles must be in general position, i.e. must not meet $\Z_2$ fixed
points, for $\kappa$
to have a well defined image in $H^2(X,\Z)$. Suppose the cycle
does meet fixed points with labels in $P\subset\F_2^4$. Then 
the $\Z_2$ quotient
produces a $2\colon\!\!1$ cover of a   sphere with branch points
$s\in P$, which on blowing up are replaced by the corresponding
exceptional divisors.
Hence $\sqrt2\kappa-\sum_{s\in P}E_s$ is Poincar\'e dual to a $2\colon\!\!1$
unbranched covering of a $K3$ cycle, i.e.
$\inv{\sqrt2}\kappa-\inv{2}\sum_{s\in P}E_s\in H^2(X,\Z)$.
Indeed, the above describes a well defined map
$$
\gamma:K_2^\ast/K_2 \longrightarrow \Pi^\ast/\Pi, \quad
\gamma(\inv{\sqrt2}\kappa):=-\inv{2}\sum_{s\in P}E_s,
$$
since for $P,\,P^\prime$ corresponding to two different non-generic
positions of our cycle, 
$\inv{2}\sum_{s\in P}E_s-\inv{2}\sum_{s\in P^\prime}E_s\in\Pi$ by \req{kummer}.

Vice versa, Theorem \ref{lattemb} together with \req{kummer} shows that
$H^2(X,\Z)$ is generated by
\begin{enumerate}
\item
$\pi_\ast (H^2(T,\Z))\cong H^2(T,\Z)(2)$,
\item
$\m E_2$, the Poincar\'e duals of the rational spheres comprising the
exceptional divisors in the blow up,
\item
forms of type $\inv{\sqrt2}\kappa-\inv{2}\sum_{s\in P}E_s\in H^2(X,\Z)$,
where $\sqrt2\kappa\in \pi_\ast H^2(T,\Z)$ determines $P$ as explained above.
\end{enumerate}
In particular, the entire lattice $H^2(X,\Z)$ is  given in terms
of two--forms that correspond to torus cycles or exceptional divisors, so for
$G=\Z_2$ the desired embedding 
$\pi_\ast(H^2(T,\Z)^G)\hookrightarrow H^2(X,\Z)$ is found.

It is obvious how to generalize i., ii. above to the other groups $G$. 
To understand forms of type iii. consider  the
following calculation  in terms of local coordinates for $G=\Z_M$:
We choose $\Z_M$ invariant polynomials $(x_1,x_2,x_3):=(z_1^M,z_2^M,z_1z_2)$
as coordinates on $T/\Z_M$ near the fixed point 
$(z_1,z_2)=(0,0)$. The blow up of $(0,0)$
is the closure of
$$
\left\{ (x=(x_1,x_2,x_3);\,s)\in\left(\C^3-\{0\}\right)\times\C\Pn^2
\mid x\sim s,\; x_1x_2=x_3^M\right\}.
$$
Near the point 
$(x;s)=(0,0,0;1,0,0)$ we use $x_1,s_3$ as coordinates and write
$(x_1,x_2,$ $x_3;s_1,s_2,s_3)
=$ $(x_1, x_1^{M-1}s_3^M, x_1s_3;1, x_1^{M-2}s_3^M, s_3)$. In this coordinate
patch, the Poincar\'e dual of $E_0$ is given by the equation $x_1=0$. 
Let $\kappa\in H^2(T,\Z)$
correspond to the cycle 
$(z_2-\zeta\eps$ for some $M^{th}$ root of unity $\zeta$, 
$\eps=const.)$, then 
its image $\sqrt M\kappa\in K_M$ corresponds to 
$(x_1^{M-1}s_3^M-\eps^M)$. So, as $\eps\rightarrow0$, our cycle
decomposes into $(M-1)(x_1) + M(s_3)$. In other words, we can 
calculate the Poincar\'e
dual $F$ of $(s_3)$ from $\sqrt M\kappa=(M-1)e+MF$, 
where $e\in\Gamma_M$
with $e=E_0+\cdots$, the dots denoting contributions from further blow ups.
This way, we can construct forms $F$ of type iii. for all relevant $G$.

We conclude that for general $G$, $H^2(X,\Z)$ contains
\renewcommand{\theenumi}{\Roman{enumi}}
\begin{enumerate}
\item
$\pi_\ast(H^2(T,\Z)^G)\cong H^2(T,\Z)^G(|G|)$,
\item
$\m E_{|G|}$, the Poincar\'e duals of the rational spheres comprising the
exceptional divisors in the blow up listed in \req{notations},
\item
forms of type $\inv{\sqrt{|G|}}\kappa+\inv{|G|}e$,
where for the dual of $\kappa\in H^2(T,\Z)^G$ we pick a non-generic position
on $T$.
Then the fixed point $s\in\m S$ occurs on that cycle with multiplicity $a_s$,
and $e=\sum_{s\in\m S}a_s E_s$ for an appropriate $E_s\in\Gamma_s$, such that  
$\inv{|\Gamma_s^\ast/\Gamma_s|}E_s\in\Gamma_s^\ast$ is primitive if
$\kappa\in H^2(T,\Z)^G$ is.
\end{enumerate}
For all but $D_4$ type
singularities $s\in\m S$,
since $\Gamma_s^\ast/\Gamma_s$ is cyclic, 
$\inv{|\Gamma_s^\ast/\Gamma_s|}E_s$   
is already  determined by III. up to a sign and contributions
from II.  In all cases,
the remaining ambiguities can be cleared up   by the fact that
$H^2(X,\Z)$ is an even lattice. Moreover, in a case by case study we find
that as in the Kummer case the vectors listed in I.-III. already generate
$H^2(X,\Z)$. Namely, 
those of type III. allow to read off generators
of $K_{|G|}^\ast/K_{|G|}$ and $K_{|G|}/\pi_\ast( H^2(T,\Z)^G)$ (or analogously
for $\Pi_{|G|}$); this turns out to determine $K_{|G|}$ already, thus
$|K_{|G|}^\ast/K_{|G|}|$ is known. Now one uses
\begin{equation}\label{discriminant}
|K_{|G|}^\ast/K_{|G|}|
= |\Pi_{|G|}^\ast/\Pi_{|G|}|
= { \disc\Gamma_{|G|}\over \left[ \Pi_{|G|}\colon \Gamma_{|G|} \right]^2}
\end{equation}
to check that all generators of $\Pi_{|G|}/\Gamma_{|G|}$ have been found.

To illustrate the above recipe we present the case $G=\Z_4$;
see Prop.\ \ref{kummerlattice} for notations.
Pick generators $\{\mu_1,\mu_2\}$, $\{\mu_3,\mu_4\}$  of
$H^1(T^2,\Z)$, $H^1(\wt T^2,\Z)$ with $\mu_i=dx_i$ with respect to 
coordinates $z_1=x_1+ix_2,\, z_2=x_3+ix_4$. 
Then $\mu_1\wedge\mu_2$ is Poincar\'e dual
to $(z_1-const.)$ and for non-generic $const.$ may
contain the fixed points $\{ (i,0,0)$, $ (i,1,0)$, $(i,0,1)$, $(i,1,1)\}$ with
 $i\in\F_2^2$. Since for $i=(0,0)$ and $i=(1,1)$ the $\Z_4$ action identifies
$(i,1,0)$ with $(i,0,1)$, we find the following
lattice vectors of type III.
$$
\inv[\sqrt4]{4} \mu_1\wedge\mu_2 + \inv{4}( 2E_{(0,0,1,0)+\eps(1,1,0,0)} 
+ E_{(0,0,0,0)+\eps(1,1,0,0)} + E_{(0,0,1,1)+\eps(1,1,0,0)}), 
$$
$\eps\in\{0,1\}.$
The cycles with $i=(1,0)$ and $i=(0,1)$ must be added to be
$\Z_4$ invariant and then give
$$
\inv[2\sqrt4]{4} \mu_1\wedge\mu_2 
+ \inv{2}( E_{(1,0,0,0)} + E_{(1,0,1,0)} + E_{(1,0,0,1)} + E_{(1,0,1,1)}).
$$
The latter vector is spurious as can be seen from the list in 
Prop.\ \ref{kummerlattice}. The other elements of $M_4$ in that proposition
are obtained analogously from cycles $(z_2-const.)$,
$(\xi_++\qu\xi_--const.)+(\xi_+-\qu\xi_--const.)$ where
$\xi_\pm:=z_1\pm iz_2$, and with $\eta_\pm:=z_1\pm z_2$ from
$(\eta_++\qu\eta_--const.)+(\eta_+-\qu\eta_--const.)$. The relative signs
of the $E_i$ are determined by the fact that $H^2(X,\Z)$ 
is an even lattice.

So far, we have found a set $M_4\subset H^2(X,\Z)$ from which we can read
generators of $\Pi_4/\Gamma_4$ as listed in Prop.\ \ref{kummerlattice}. 
Moreover, we find
$\mu_1\wedge\mu_3+\mu_4\wedge\mu_2$, 
$\mu_1\wedge\mu_4+\mu_2\wedge\mu_3\in K_4$.
We remark that these forms can be used to generate $H^{2,0}(T,\Z)$
and $H^{0,2}(T,\Z)$ for $T=\R^4/\Z^4$, and by the above observation
the transcendental lattice of the corresponding $\Z_4$ orbifold has quadratic
form $\diag(2,2)$. This is in
agreement with \cite[Lemma 5.2]{shin77}. 

As to the construction of
$H^2(X,\Z)$, since from $M_4$ we find 
$\inv{2}\mu_1\wedge\mu_2,\inv{2}\mu_3\wedge\mu_4$,
$\inv{2}(\mu_1\wedge\mu_3+\mu_4\wedge\mu_2)$,
$\inv{2}(\mu_1\wedge\mu_4+\mu_2\wedge\mu_3)\in K_4^\ast$,
we conclude
$$
K_4=\spann_\Z( 2\mu_1\wedge\mu_2, 2\mu_3\wedge\mu_4,
\mu_1\wedge\mu_3+\mu_4\wedge\mu_2, \mu_1\wedge\mu_4+\mu_2\wedge\mu_3),
$$
and $|K_4^\ast/K_4|=4^3$. Hence \req{discriminant} shows
$[\Pi_4\colon \Gamma_4]=16$ and proves that the three vectors listed in 
Prop.\ \ref{kummerlattice}  generate 
$\Pi_4/\Gamma_4\cong\Z_4\times\Z_2^2$ and therefore together
with $\m E_4$ suffice to generate $\Pi_4$.
\begin{envis}{Proposition}{kummerlattice}
Let $X$ denote an orbifold limit \smash{$\wt{T/G}$} of $K3$, 
where $G$ does not contain non-trivial
translations. In other words \cite{fu88},
$G=\Z_M$ with
$M\in\{2,3,4,6\}$, $G=\wh{D}_n$, $n\in\{4,5\}$, $G=\wh D_4^\prime$, or
$G=\wh\T$. Then 
$H^2(X,\Z)$ is generated by 
$\pi_\ast (H^2(T,\Z)^G)\cong H^2(T,\Z)^G(|G|)$, the Poincar\'e duals
$\m E_{|G|}$ of components of
exceptional divisors listed in \req{notations}, and
the set $M_{|G|}$ given by iii. for $G=\Z_2$ or otherwise listed below.

Here,   
for each type of fixed point $s\in\m S$  
we fix   generators
$\inv{|\Gamma_s^\ast/\Gamma_s|} E_s$ 
or $\inv{2} E_s^{(a,b)}$
of $\Gamma^\ast_s/\Gamma_s$ with
$$
\begin{array}{lrcl}\ds
t\in\F_3^2:
\>E_t\>\!\!:=\!\!\>E_t^{(1)}+2E_t^{(2)}; 
\; i\in I^{(4)}\!\!:\,E_i:=E_i^{(1)}+2E_i^{(2)}+3E_i^{(3)};\e
G=\Z_6:\>E_0^{\Z_6}
\>\!\!:=\!\!\>E_0^{(1)}+2E_0^{(2)}+3E_0^{(3)}+4E_0^{(4)}+5E_0^{(5)} ; \e
G=\wh D_4^{(\prime)}\!\!:\!\!\!\!
\>E_i^{(a,b)} \!\!
\>\!\!:=\!\!\> E_i^{(a)}+E_i^{(b)}\;\;\;
(i\in \{0,1\} \mbox{ or } i\in I^{(4)};\, a,b\in\{1,2,3\});\e
G=\wh D_5:\>E_0^{\wh D_5}
\>\!\!:=\!\!\>5E_0^{(3)}+6E_0^{(5)}+3E_0^{(4)}+4E_0^{(1)}+2E_0^{(2)}; \e
G=\wh\T:\>E_0^{E_6}
\>\!\!:=\!\!\> E_0^{(1)}+2E_0^{(2)}+4E_0^{(4)}+5E_0^{(5)}.
\end{array}
$$
With standard basis
$\{ f_j\}$ of $\,\F_2^4$ let $P_{jk}:=\spann_{\F_2}\{f_j,f_k\}$.
Moreover, $\{\mu_j,$ $j\in\{1,\dots,4\}\}$ 
always denotes an appropriate basis of
$H^1(T,\Z)$ such that
$\mu_j\wedge\mu_k,\, j,k\in\{1,\dots,4\}$    generate $H^2(T,\Z)$.
More precisely, if $\Z_4\subset G$, then a generator $\zeta\in\Z_4$
acts by 
$$
\zeta: \;(\mu_1,\,\mu_2,\,\mu_3,\,\mu_4) \longmapsto
(\mu_2,\,-\mu_1,\,-\mu_4,\,\mu_3),
$$
and for $G=\Z_3,\,\Z_6$ and $\wh D_5$ a $\Z_3$ generator $\zeta^\prime$
acts by
$$
\zeta^\prime: \;(\mu_1,\,\mu_2,\,\mu_3,\,\mu_4) \longmapsto
(\mu_2-\mu_1,\,-\mu_1,\,-\mu_4,\,\mu_3-\mu_4).
$$
We find:
$$
M_3=\left\{
\begin{array}{l}\ds
\inv{\sqrt3} \mu_1\wedge\mu_2 + \inv{3}\left( E_{(i,0)}+E_{(i,1)}+E_{(i,2)} \right), 
\quad i\in\F_3,\ed
\inv{\sqrt3} \mu_3\wedge\mu_4 - \inv{3}\left( E_{(0,i)}+E_{(1,i)}+E_{(2,i)} \right), \quad i\in\F_3,\ed
\inv{\sqrt3} (\mu_1-\mu_3)\wedge(\mu_2-\inv{2}\mu_1+\mu_4-\inv{2}\mu_3)\ed
\hphantom{\inv{\sqrt3} (\mu_1-\mu_4)\wedge(\mu_2-\mu_3)}
+ \inv{3}\left( E_{(0,0)}+E_{(1,2)}+E_{(2,1)} \right), \ed
\inv{\sqrt3} (\mu_1-\mu_4)\wedge(\mu_2-\mu_3)
+ \inv{3}\left( E_{(0,0)}+E_{(1,1)}+E_{(2,2)} \right) 
\end{array}
\right\},
$$
generators of $\Pi_3/\Gamma_3$:
$$
\inv{3}\left( \sum_{t\in L}E_t - \sum_{t^\prime\in L^\prime}E_{t^\prime}\right),
\quad L,L^\prime\subset\F_3^2 \mbox{ parallel lines}.
$$
\myline
$$
M_4=\left\{
\begin{array}{l}\ds
 \inv{2}\mu_1\wedge\mu_2+ \inv{2} E_{(0,0,1,0)+\eps(1,1,0,0)}\ed
\hphantom{(\mu_2-\inv{2}\mu_1+\mu_4-\inv{2}\mu_3)}
+ \inv{4} \sum_{i\in P_{34}\cap I^{(4)}} E_{i+\eps(1,1,0,0)},
\eps\in\{0,1\},\edd
\inv{2}\mu_3\wedge\mu_4- \inv{2}E_{(1,0,0,0)+\eps(0,0,1,1)}\ed
\hphantom{(\mu_2-\inv{2}\mu_1+\mu_4-\inv{2}\mu_3)}
- \inv{4} \sum_{i\in P_{12}\cap I^{(4)}} E_{i+\eps(0,0,1,1)}
,\eps\in\{0,1\}, \edd
\inv{2}\left(\mu_1\wedge\mu_3+\mu_4\wedge\mu_2\right)
- \inv{2}\sum_{i\in P_{13}} E_{i+j}+E_j,\quad j\in I^{(4)} ,\edd
\inv{2}\left(\mu_1\wedge\mu_4+\mu_2\wedge\mu_3\right)
- \inv{2}\sum_{i\in P_{14}} E_{i+j}+E_j,\quad j\in I^{(4)} 
\end{array}
\right\},
$$
generators of $\Pi_4/\Gamma_4$:
$$
\begin{array}{l}\ds
\inv{4}\left(
E_{(0,0,0,0)} + E_{(1,1,0,0)} - E_{(0,0,1,1)} - E_{(1,1,1,1)}\right) \ed
\hphantom{\inv{2}\left(E_{(0,0,0,0)} +  E_{(1,0,0,0)} 
+ E_{(1,0,1,0)}\right.\!}
+\inv{2}\left( E_{(1,0,0,0)} - E_{(1,0,1,1)} \right),\edd
\inv{2}\left(E_{(0,0,0,0)} +  E_{(1,0,0,0)} 
+ E_{(1,0,1,0)} + E_{(1,0,0,1)} - E_{(0,0,1,1)} + E_{(1,0,1,1)} \right) ,\edd
\inv{2}\left(E_{(0,0,0,0)}  + E_{(0,0,1,0)} 
+ E_{(1,0,1,0)} + E_{(1,0,0,1)} -  E_{(1,1,0,0)}+ E_{(1,1,1,0)}\right) 
.
\end{array}
$$
\myline
$$
M_6=\left\{
\begin{array}{l}\ds
\inv{\sqrt6}\mu_1\wedge\mu_2 + \inv{6}E_0^{\Z_6}+\inv{3} E_{(0,1)}
+\inv{2} E_{(0,0,0,1)},\ed
\inv{\sqrt6}\mu_3\wedge\mu_4 - \inv{6}E_0^{\Z_6}-\inv{3} E_{(1,0)}
-\inv{2} E_{(1,0,0,0)}, 
\ed
\inv[2]{\sqrt6}\mu_1\wedge\mu_2 
+ \!\inv{3}\left( E_{(1,0)}+E_{(1,1)}+E_{(1,2)}\right),\ed
\inv[2]{\sqrt6}\mu_3\wedge\mu_4 - \!\inv{3}\left( E_{(0,1)}+E_{(1,1)}
+E_{(1,2)}\right),\ed 
\inv[3]{\sqrt6}\mu_1\wedge\mu_2 
+ \inv{2}\left( E_{(1,0,0,0)}+E_{(1,0,1,0)}+E_{(1,0,0,1)}+E_{(0,1,0,1)}\right),\ed
\inv[3]{\sqrt6}\mu_3\wedge\mu_4 
- \inv{2}\left( E_{(0,0,0,1)}+E_{(1,0,1,0)}+E_{(1,0,0,1)}+E_{(0,1,0,1)}\right),\ed
\inv{\sqrt6}(\mu_1-\mu_3)\wedge(\mu_2-\inv{2}\mu_1+\mu_4-\inv{2}\mu_3)\ed
\hphantom{\inv{\sqrt6}(\mu_1-\mu_4)\wedge(\mu_2-\mu_3)}
+\inv{6}E_0^{\Z_6}+\inv{3}E_{(1,2)}+\inv{2} E_{(1,0,0,1)} ,\ed
\inv{\sqrt6}(\mu_1-\mu_4)\wedge(\mu_2-\mu_3)
+\inv{6}E_0^{\Z_6}+\inv{3}E_{(1,1)}+\inv{2} E_{(0,1,0,1)} 
\end{array}\right\},
$$
generator of $\Pi_6/\Gamma_6$:
$$
\begin{array}{l}
\inv{6} E_0^{\Z_6}
+ \inv{3}\left( E_{(1,0)}+E_{(0,1)}+E_{(1,1)}+E_{(1,2)}\right)\ed
\hphantom{\inv{6} E_0^{\Z_6}}
+ \inv{2}\left( E_{(1,0,0,0)}+E_{(0,0,0,1)}+E_{(1,0,1,0)}+E_{(1,0,0,1)}+E_{(0,1,0,1)}\right).
\end{array}
$$
\myline
$$
M_8=\left\{
\begin{array}{l}\ds
\inv{\sqrt8}(\mu_1\wedge\mu_2+\mu_3\wedge\mu_4)
- \inv{2}E_{(1,0,0,0)} - \inv{4}E_{(1,1,0,0)}
- \inv{2}   E_0^{(1,2)}  , \ed
\inv{\sqrt8}(\mu_1\wedge\mu_2+\mu_3\wedge\mu_4)
+ \inv{2}E_{(0,1,1,1)} + \inv{4}E_{(1,1,0,0)}
+ \inv{2}  E_1^{(1,2)}   , \ed
\inv[2]{\sqrt8}(\mu_1\wedge\mu_2+\mu_3\wedge\mu_4)
- \inv{2}\sum_{i\in\F_2^2} E_{(1,0,i)}, \ed
\inv{\sqrt8}(\mu_1\wedge\mu_3+\mu_4\wedge\mu_2)
- \inv{2}E_{(1,0,0,0)} - \inv{4}E_{(1,0,1,0)}
- \inv{2}   E_0^{(1,3)}  , \ed
\inv{\sqrt8}(\mu_1\wedge\mu_3+\mu_4\wedge\mu_2)
+ \inv{2}E_{(0,1,1,1)} + \inv{4}E_{(1,0,1,0)}
+ \inv{2}   E_1^{(1,3)}  , \ed
\inv[2]{\sqrt8}(\mu_1\wedge\mu_3+\mu_4\wedge\mu_2)
- \inv{2}\sum_{i_1,i_2\in\F_2} E_{(i_1,1,i_2,0)}, \ed
\inv{\sqrt8}(\mu_1\wedge\mu_4+\mu_2\wedge\mu_3)
- \inv{2}E_{(1,0,0,0)} - \inv{4}E_{(1,0,0,1)}
- \inv{2}   E_0^{(2,3)}  , \ed
\inv{\sqrt8}(\mu_1\wedge\mu_4+\mu_2\wedge\mu_3)
+ \inv{2}E_{(0,1,1,1)} + \inv{4}E_{(1,0,0,1)}
+ \inv{2}  E_1^{(2,3)}  , \ed
\inv[2]{\sqrt8}(\mu_1\wedge\mu_4+\mu_2\wedge\mu_3)
- \inv{2}\sum_{i\in\F_2^2} E_{(1,i,0)}
\end{array}
\right\},
$$
generators of $\Pi_8/\Gamma_8$:
$$
\begin{array}{l}\ds
\inv{2}\left(
E_{(1,0,0,0)} + E_{(0,1,1,1)} + E_{(1,1,0,0)}
+E_0^{(1,2)}+E_1^{(1,2)}
\right) ,\ed
\inv{2}\left(
E_{(1,0,0,0)} + E_{(0,1,1,1)} + E_{(1,0,1,0)}
+E_0^{(1,3)}+E_1^{(1,3)}
\right) ,\ed
\inv{2}\left(
E_{(1,0,0,0)} + E_{(0,1,1,1)} + E_{(1,0,0,1)}
+E_0^{(2,3)}+E_1^{(2,3)}
\right) .
\end{array}
$$
\myline
$$
M_8^\prime=\left\{
\begin{array}{l}\ds
\inv{\sqrt8}(\mu_1\wedge\mu_3+\mu_4\wedge\mu_2+
\mu_1\wedge\mu_4+\mu_2\wedge\mu_3 +2\mu_4\wedge\mu_3) \ed
\hphantom{\mu_1\wedge\mu_4+\mu_2\wedge\mu_3}
+ \inv{2} \left( E_{(0,0,0,0)}^{(1,2)} + E_{(1,1,0,0)}^{(1,2)} 
+ E_{(0,0,0,1)}
\right),\ed
\inv{\sqrt8}(\mu_1\wedge\mu_3+\mu_4\wedge\mu_2+
\mu_1\wedge\mu_4+\mu_2\wedge\mu_3 +2\mu_4\wedge\mu_3)\ed
\quad
+ \inv{2} \left( E_{(0,0,0,0)}^{(1,2)} 
+ E_{(1,1,0,0)}^{(2,3)} 
+ E_{(0,0,1,1)}^{(2,3)} 
+ E_{(1,1,1,1)}^{(1,2)} 
+ E_{(0,0,0,1)}
\right),\ed
\inv{\sqrt8}(\mu_1\wedge\mu_3+\mu_4\wedge\mu_2+
\mu_1\wedge\mu_4+\mu_2\wedge\mu_3 +2\mu_4\wedge\mu_3)\ed
\quad
+ \inv{2} \left( E_{(0,0,0,0)}^{(1,2)} 
+ E_{(1,1,0,0)}^{(1,3)} 
+ E_{(0,0,1,1)}^{(1,2)} 
+ E_{(1,1,1,1)}^{(1,3)} 
+ E_{(0,0,0,1)}
\right),\ed
\inv[2]{\sqrt8}(\mu_1\wedge\mu_2+\mu_3\wedge\mu_4)
+ \inv{2}\left( E_{(1,0,0,0)} +  E_{(0,0,0,1)} \right), \ed
\inv[2]{\sqrt8}(\mu_1\wedge\mu_3+\mu_4\wedge\mu_2+\mu_4\wedge\mu_3)
+ \inv{2}\left( E_{(1,0,1,0)} +  E_{(0,0,0,1)} \right), \ed
\inv[2]{\sqrt8}(\mu_1\wedge\mu_3+\mu_4\wedge\mu_2+\mu_4\wedge\mu_3)
+ \inv{2}\left( E_{(1,0,1,0)} +  E_{(0,0,0,1)} \right)\ed
\hphantom{\mu_1\wedge\mu_4+\mu_2\wedge\mu_3} 
+ \inv{2}\sum_{i\in I^{(4)}}
E_i^{(a,b)}, \quad\quad
a,b\in\{1,2,3\}, a\neq b, \ed
\inv[2]{\sqrt8}(\mu_1\wedge\mu_4+\mu_2\wedge\mu_3+\mu_4\wedge\mu_3)
+ \inv{2}\left( E_{(1,0,1,0)} +  E_{(0,0,0,1)} \right), \ed
\inv[2]{\sqrt8}(\mu_1\wedge\mu_4+\mu_2\wedge\mu_3+\mu_4\wedge\mu_3)
+ \inv{2}\left( E_{(1,0,1,0)} +  E_{(0,0,0,1)} \right)\ed
\hphantom{\mu_1\wedge\mu_4+\mu_2\wedge\mu_3} 
+ \inv{2}\sum_{i\in I^{(4)}}E_i^{(a,b)},
\quad\quad
a,b\in\{1,2,3\}, a\neq b 
\end{array}
\right\},
$$
generators of $\Pi_8^\prime/\Gamma_8^\prime$:
$$
\begin{array}{c}\ds
\inv{2}\left(
E_{(1,1,0,0)}^{(1,3)} + E_{(0,0,1,1)}^{(2,3)} + E_{(1,1,1,1)}^{(1,2)}\right),
\;\;
\inv{2}\left(
E_{(1,1,0,0)}^{(2,3)} + E_{(0,0,1,1)}^{(1,2)} + E_{(1,1,1,1)}^{(1,3)}\right), 
\ed
\inv{2}\sum_{i\in I^{(4)}}
E_i^{(a,b)},\quad a,b\in\{1,2,3\},\; a\neq b .
\end{array}
$$
\myline
$$
M_{12}=\left\{
\begin{array}{l}\ds
\inv{\sqrt{12}}(\mu_1\wedge\mu_2+\mu_3\wedge\mu_4)
- \inv{2} E_0^{\wh D_5}- \inv{3}E_{(1,0)} - \inv{2}E_{(1,0,0,0)} , \ed
\inv[3]{\sqrt{12}}(\mu_1\wedge\mu_2+\mu_3\wedge\mu_4)
- \inv{2}\sum_{i\in\F_2^2} E_{(1,0,i)} ,\ed
\inv[2]{\sqrt{12}}(\mu_1\wedge\mu_2+\mu_3\wedge\mu_4)
+ \inv{3} E_{(1,0)}, \ed
\inv[2]{\sqrt{12}}(\mu_1\wedge\mu_4+\mu_2\wedge\mu_3+\mu_3\wedge\mu_1) \ed
\hphantom{\inv[2]{\sqrt{12}}(\mu_1\wedge\mu_3+\mu_4\wedge\mu_2)}
- \inv{2} E_0^{\wh D_5}- \inv{3}E_{(1,1)} 
- \inv{2} E_{(1,0,1,0)} , \ed
\inv[2]{\sqrt{12}}(\mu_1\wedge\mu_3+\mu_4\wedge\mu_2)
- \inv{2} E_0^{\wh D_5}- \inv{3}E_{(1,1)} - \inv{2}E_{(1,0,0,1)}
\end{array}
\right\},
$$
generator of $\Pi_{12}/\Gamma_{12}$:
$$
\inv{2}\left( E_0^{\wh D_5} +
 E_{(1,0,1,0)} +E_{(1,0,0,1)} + E_{(0,1,0,1)}
\right) .
$$
\myline
$$
M_{24}=\left\{
\begin{array}{l}\ds 
\inv[3]{\sqrt{24}}\!(\mu_1\wedge\mu_3+\mu_4\wedge\mu_2+
\mu_1\wedge\mu_4+\mu_2\wedge\mu_3 +2\mu_4\wedge\mu_3)  \!\ed
\hphantom{mu_1\wedge\mu_3+\mu_4\wedge\mu_2}
+\! \inv{2}  \!\left( E_{(1,1,0,0)}^{(1,2)} + E_{(0,0,0,1)} \right)\!,\ed  
\inv[6]{\sqrt{24}}(\mu_1\wedge\mu_2+\mu_3\wedge\mu_4)
+ \inv{2} E_{(1,1,0,0)}^{(1,3)} , \ed
\inv[6]{\sqrt{24}}(\mu_1\wedge\mu_3+\mu_4\wedge\mu_2+\mu_4\wedge\mu_3)\ed
\hphantom{mu_1\wedge\mu_3+\mu_4\wedge\mu_2}
+ \inv{3} E_0^{E_6} + \inv{3} E_{(1,1)} + \inv{2} E_{(1,1,0,0)}^{(2,3)} , \ed
\inv[6]{\sqrt{24}}(\mu_1\wedge\mu_3+\mu_4\wedge\mu_2+\mu_4\wedge\mu_3)\ed
\hphantom{mu_1\wedge\mu_3+\mu_4\wedge\mu_2}
- \inv{3}\left( E_{(1,0)}+E_{(0,1)}+E_{(1,2)}\right)  
+ \inv{2} E_{(1,1,0,0)}^{(2,3)} , \ed
\inv[6]{\sqrt{24}}(\mu_1\wedge\mu_4+\mu_2\wedge\mu_3+\mu_4\wedge\mu_3)\ed
\hphantom{mu_1\wedge\mu_3+\mu_4\wedge\mu_2}
+ \inv{3} E_0^{E_6} + \inv{3} E_{(1,1)} + \inv{2} E_{(1,1,0,0)}^{(2,3)} , \ed
\inv[6]{\sqrt{24}}(\mu_1\wedge\mu_4+\mu_2\wedge\mu_3+\mu_4\wedge\mu_3)\ed
\hphantom{mu_1\wedge\mu_3+\mu_4\wedge\mu_2}
- \inv{3}\left( E_{(1,0)}+E_{(0,1)}+E_{(1,2)} \right)
+ \inv{2} E_{(1,1,0,0)}^{(2,3)} , \ed
\end{array}
\right\},
$$
generator of $\Pi_{24}/\Gamma_{24}$:
$$
\inv{3} E_0^{E_6}
+ \inv{3}\left( E_{(1,1)}+E_{(1,0)}+E_{(0,1)}+E_{(1,2)}\right) . 
$$
\myline
\end{envis}\indent
Prop.\ \ref{kummerlattice} pinpoints the characteristic
distinction between our discussion of compact orbifolds as opposed to the
approach of \cite{do97,blin97}. On $Y=\wt{\C^2/G}$, the components of 
exceptional divisors generate $H_2(Y,\Z)$, whereas on our 
$K3$ surface $X$ it is a nontrivial
problem to determine the primitive sublattices of $H_2(X,\Z)$ that contain
the exceptional divisors.
\section{Consistent embedding of orbifold conformal field theories in 
$\cal M^{K3}$}\label{quantum}
Prop.\ \ref{kummerlattice} serves to explicitly
locate $G$ orbifold limits of  Einstein metrics of volume $1$ on $K3$ 
 within the moduli space 
$$O^+(H^2(X,\Z))\backslash O^+(H^2(X,\R)) / SO(3)\times O(19)$$ of
such metrics. Analogous reasoning should enable us to locate $G$ orbifold
conformal field theories in the moduli space
$$O^+(H^{even}(X,\Z))\backslash O^+(H^{even}(X,\R)) / SO(4)\times O(20)$$ of
conformal field theories associated to $K3$. 
Again, the construction of the appropriate
embedding $\wh\pi_\ast (H^{even}(T,\Z)^G)\hookrightarrow H^{even}(X,\Z)$
is all we need, where $\wh\pi_\ast$ is an extension of $\pi_\ast$. Hence 
we must
require the image $x\subset H^{even}(X,\R)$ of a four--plane 
$x_T\in\m T^{4,4}$, which describes
 a toroidal conformal field theory with $G$ symmetry, to admit
a geometric interpretation on the corresponding $G$ orbifold limit of $K3$.
This statement is made precise by the use of  \req{asmois}, which
assigns a geometric interpretation to any of our conformal field 
theories: 

Suppose 
$\upsilon,\upsilon^0\in H^{even}(T,\Z)$ are primitive null vectors
with $\skp{\upsilon}{\upsilon^0}=1$ such that $x_T$ has geometric 
interpretation $(\Sigma_T,V_T,B_T)$. We need to find primitive null vectors
$\wh\upsilon,\wh\upsilon^0\in H^{even}(X,\Z)$ with 
$\skp{\wh\upsilon}{\wh\upsilon^0}=1$ such that $x$ has geometric interpretation
$(\Sigma,V,B)$, where $\Sigma=\pi_\ast\Sigma_T$. The location of this 
three--plane in $H^2(X,\R)$ is fixed by Prop.\ \ref{kummerlattice}.
In particular, for each $G$ orbifold,
the two--forms corresponding to exceptional divisors
of the blow up must be contained in $H^{even}(X,\Z)$ in such a way that
the exceptional divisors have volume zero:
\begin{equation}\label{orbint}
\spann_\R(K_{|G|} ,\wh\upsilon,\wh\upsilon^0)^\perp \cap H^{even}(X,\Z)
\supset \wh\Pi_{|G|} \cong\Pi_{|G|}.
\end{equation}
Let $\Lambda_{|G|}$ denote the primitive sublattice of $H^{even}(X,\Z)$
which contains the lattice
$\wh\pi_\ast (H^{even}(T,\Z)^G)$. Then  
$\Lambda_{|G|}\cong K_{|G|}\oplus U(|G|)$, where $U(|G|)$ is generated by
the $\wh\pi_\ast$ images $\sqrt{|G|}\upsilon, \sqrt{|G|}\upsilon^0$
of $\upsilon, \upsilon^0$. Any ansatz with non--primitive 
$\sqrt{|G|}\upsilon$ or $\sqrt{|G|}\upsilon^0$ leads to contradictions
by the methods presented below, cf. our comment below Theorem \ref{z4emb}. 
Here, $\sqrt{|G|}\upsilon$ is the
Poincar\'e dual of a generic point on $X$, and 
$\sqrt{|G|}\upsilon^0$ denotes the dual of the cycle obtained as
closure of $\pi(T-\wt{\m S})$ on $X$; this interpretation is in accord with
$\skp{\sqrt{|G|}\upsilon}{\sqrt{|G|}\upsilon^0}=|G|$, since $\pi$ is 
$|G|\colon\!\!1$ outside the set $\m S$ of fixed points. 
The ad hoc assignment of equal scaling factors $\sqrt{|G|}$
to both $\upsilon$ and
$\upsilon^0$ is chosen for ease of notation; this freedom of choice drops
out in all results below, see \req{bfield}. 

It follows that we cannot use
$\sqrt{|G|}\upsilon, \sqrt{|G|}\upsilon^0$ for 
$\wh\upsilon, \wh\upsilon^0$:
\begin{equation}\label{quafo}
(\Lambda_{|G|})^\ast/\Lambda_{|G|}
\;\cong\; (K_{|G|})^\ast/K_{|G|} \times \Z_{|G|}^2
\;\cong\; (\Pi_{|G|})^\ast/\Pi_{|G|} \times \Z_{|G|}^2
\end{equation}
by Theorem \ref{lattemb}, so again by Theorem \ref{lattemb},
$(\Lambda_{|G|})^\perp\cap H^{even}(X,\Z)\not\cong\Pi_{|G|}$.

Since $\sqrt{|G|}\upsilon$ has a good geometric interpretation
as Poincar\'e dual of the generic point on $X$, we use the ansatz
\begin{equation}\label{newmlattice}
\wh\upsilon:=\sqrt{|G|}\,\upsilon, \quad
\wh\upsilon^0 := \inv{\sqrt{|G|}}\upsilon^0 - \inv{|G|} B_{|G|}
- \inv[\|B_{|G|}\|^2]{2|G|^2} \sqrt{|G|}\upsilon,
\end{equation}
with $B_{|G|}\perp\upsilon,\upsilon^0$ to be determined. 
By \cite{lope81,ni80} the automorphism
group $O^+(H^{even}(X,\Z))$
in \req{modsp} acts transitively on pairs of primitive lattice vectors of
equal length. Hence \req{newmlattice} is also the most general ansatz we need.
Assume that 
for given $G$ we have
found $B_{|G|}$ such that \req{orbint} holds for
$\wh\upsilon,\wh\upsilon^0$ as in \req{newmlattice} (we will show that 
$B_{|G|}$ is uniquely determined up to lattice automorphisms). All calculations
below are carried out in $H^{even}(X,\Q)$. By 
$Y:H^{even}(X,\Q)\longrightarrow H^{even}(X,\Q)$ we denote the orthogonal
projection onto $\upsilon^\perp\cap(\upsilon^0)^\perp$ and (by a slight
abuse of notation; see the footnote on page \pageref{abuse}) set
$\Pi_{|G|}:=Y(\wh\Pi_{|G|})$, because indeed $\Pi_{|G|}\cong\wh\Pi_{|G|}$
since $\wh\Pi_{|G|}\perp\upsilon$.
Then
\begin{envis}{Lemma}{correctansatz}
Suppose that $\wh\upsilon,\wh\upsilon^0\in H^{even}(X,\Z)$ have the form
\req{newmlattice} and  that for every $G$ orbifold conformal field theory
determined by
$x=\wh\pi_\ast x_T$ they give a geometric interpretation on the 
corresponding $G$ orbifold limit of $K3$ with \req{orbint}. Then
\begin{eqnarray*}
B_{|G|}\in \Pi_{|G|}, \quad \|\inv{|G|} B_{|G|}\|^2 &\in&2\Z,\\
\skp{B_{|G|}}{E}&\equiv&-1\mod {|G|} \mbox{ for some } 
E\in \Pi_{|G|}.
\end{eqnarray*}
Set $\wh M_{|G|}:=M_{|G|}\cup\{\wh\upsilon,\wh\upsilon^0\}$ with $M_{|G|}$
as defined in Prop.\ \ref{kummerlattice}, and
$$
\fa E\in \Pi_{|G|}:\quad 
\wh E:=E-\langle E,\wh\upsilon^0\rangle\,\wh\upsilon 
= E+\inv{\sqrt{|G|}} \langle B_{|G|}, E\rangle\,\upsilon.
$$
Then $\wh M_{|G|}$ and 
$\wh{\m E}_{|G|}:=\{\wh E\mid E\in \Pi_{|G|}\}$
generate  $H^{even}(X,\Z)\cong\Gamma^{4,20}$.
\end{envis}\indent
\bpr
By Theorem \ref{lattemb}, we need
to find $P_{|G|}:=(\Lambda_{|G|})^\perp\cap H^{even}(X,\Z)$, then 
 $\Lambda_{|G|}^\ast/\Lambda_{|G|}\cong P_{|G|}^\ast/P_{|G|}$,
with isomorphism denoted by $\gamma$, and the discriminant
forms agree up to a sign. This will give
\begin{equation}\label{bigglue}
H^{even}(X,\Z)\cong\left\{ (x,y)\in\Lambda_{|G|}^\ast\oplus P_{|G|}^\ast \left|
\gamma(\qu x)=\qu y\right.\right\}.
\end{equation}
We claim that $P_{|G|}=\wh P_{|G|}$ with
$$
\wh P_{|G|}:=\left\{ p\in \Pi_{|G|} \mid \skp{p}{\wh\upsilon^0}\in\Z\right\}.
$$
Namely, 
for $p\in\wh P_{|G|}$ by construction we can find $\wt p\in H^{even}(X,\Z)$
such that $\wt p-p=a\upsilon,\, a\in\R$. Since 
$
\Z\ni\langle\wt p,\wh\upsilon^0\rangle
= \inv[a]{\sqrt{|G|}} + \langle p,\wh\upsilon^0\rangle,
$
$a$ is an integral multiple of $\sqrt{|G|}$, and therefore 
$\wh P_{|G|}\subset H^{even}(X,\Z)$. But
$P_{|G|}\otimes\R=\wh P_{|G|}\otimes\R$ is clear from $P_{|G|}\subset\Pi_{|G|}$
on dimensional grounds, so
$P_{|G|}=\wh P_{|G|}$ since both are primitive sublattices of $H^{even}(X,\Z)$
by construction.

From Theorem \ref{lattemb} we conclude that $B_{|G|}$ must be chosen such
that $\wh P_{|G|}^\ast/\wh P_{|G|}\cong\Lambda_{|G|}^\ast/\Lambda_{|G|}$ with 
discriminant forms of opposite sign. Because
$\wh P_{|G|}\subset \Pi_{|G|}\subset \Pi_{|G|}^\ast\subset\wh P_{|G|}^\ast$, we can use
the decomposition
\begin{equation}\label{decomp}
\wh P_{|G|}^\ast/\wh P_{|G|} 
\cong\wh P_{|G|}^\ast/\Pi_{|G|}^\ast\times \Pi_{|G|}^\ast/\Pi_{|G|}\times \Pi_{|G|}/\wh P_{|G|} ,
\end{equation}
so from \req{quafo} we deduce
$$
\Pi_{|G|}/\wh P_{|G|}\cong \wh P_{|G|}^\ast/\Pi_{|G|}^\ast\cong \Z_{|G|}.
$$
Moreover, ${1\over {|G|}}B_{|G|}$ generates $\wh P_{|G|}^\ast/\Pi_{|G|}^\ast$,
thus $B_{|G|}\in \Pi_{|G|}^\ast$. Since the quadratic forms of
$\wh P_{|G|}^\ast/\wh P_{|G|}$ and $\Lambda_{|G|}^\ast/\Lambda_{|G|}$ agree
up to a sign as forms with values in $\Q/2\Z$, 
we conclude $\|\inv{|G|}B_{|G|}\|^2\in2\Z$, and by \req{decomp} there exists  
$E\in \Pi_{|G|}$ which generates $\Pi_{|G|}/\wh P_{|G|}$ such that
$\skp{B_{|G|}}{E}\equiv-1\mod {|G|}$. Furthermore,  by \req{newmlattice}
$B_{|G|}\in P_{|G|}=\wh P_{|G|}\subset \Pi_{|G|}$.
The generators of $H^{even}(X,\Z)$ can now be read off from \req{bigglue}
and Prop.\ \ref{kummerlattice}.
\epr
The properties listed  in Lemma \ref{correctansatz}
do not  determine $B_{|G|}$  in \req{newmlattice} uni\-que\-ly.
But since 
a shift of $\wh\upsilon^0$ by an element of $H^{even}(X,\Z)$ corresponds
to an integral shift of the B--field in the geometric interpretation
and thus is irrelevant to our discussion (see \req{modsp}), 
we can restrict ourselves
to a finite number of candidates for $B_{|G|}$. 
A lot of them will be equivalent by lattice automorphisms in 
$O^+(H^2(X,\Z))$. 
The lift $\wh B_{|G|}$ of $B_{|G|}\in \Pi_{|G|}$ to 
$\wh\Pi_{|G|}$ will determine
the offset ${1\over {|G|}}\wh B_{|G|}$ of the B--field induced on the 
exceptional divisors of the blow up by the orbifold process 
(see \req{bfield}). 
Since this is a local effect for each fixed point, 
the result for $G=\Z_2$ will determine the contribution of $\Z_2$ fixed points
for all $G$ etc.

Moreover, algebraic symmetries of the
underlying toroidal conformal field theory induce symmetries of the
orbifold conformal field theory that must not be destroyed by the B--field.
In particular, $\wh B_{|G|}$ is invariant under all algebraic automorphisms
of the orbifold limit of $K3$. 
For $G=\Z_2$ and
$G=\Z_4$ we can use the results
\cite[Thms.\ 2.7, 2.12]{nawe00} on algebraic automorphisms of 
orbifold conformal field theories obtained from toroidal theories 
on $T=\R^4/\Z^4$ to verify that all $\Z_2$ and
all $\Z_4$ type fixed points are related
by symmetries, respectively,
and therefore confirm that they
must give the same contribution to $B_{|G|}$. Moreover, in the $\Z_4$ case, 
all $E_i^{(1;3)}\!$, $i\in I^{(4)}$, must carry the same B--field flux.
Analogous reasoning severely restricts the number of candidates for
$B_{|G|}$ in all cases. Actually,
\begin{envis}{Lemma}{bchoice}
For   $G\subset SU(2)$ as in \req{notations}, the vector
$B_{|G|}$ in \req{newmlattice} is uniquely fixed,
up to lattice automorphisms in $O^+(H^2(X,\Z))$ and   shifts
of $\wh\upsilon^0$ by lattice vectors,
by the properties
listed in Lemma \ref{correctansatz} and consistency with symmetries
of $G$ orbifold conformal field theories. 

For a $G^\prime\subset G$ type
fixed point $s\in\m S$ let 
$\sum_{j} n_s^{(j)} E_s^{(j)}$ 
denote the highest root in
$\Gamma_s$. Then we can characterize $B_{|G|}$ by
$$
\fa s\in\m S, \;
\fa j:\quad
{1\over |G|} \skp{B_{|G|}}{E_s^{(j)}} =  {n_s^{(j)}\over |G^\prime|}   .
$$
\vspace*{-1em}
\end{envis}\indent
\bpr
The result $B_2=-\inv{2} \sum_{i\in\F_2^4} E_i$
for $G=\Z_2$ follows immediately from Lemma \ref{correctansatz}
together with the observation that all fixed points contribute equally.

We only add the proof for $G=\Z_4$, since the other cases are obtained
analogously.
The most general ansatz for $B_4\in\Pi_4$ that is consistent
with the symmetries of the $\Z_4$ orbifold of the toroidal model on
$\R^4/\Z^4$ and our knowledge of the 
B--field over the $\Z_2$ fixed points is 
$$
B_4 
=-\sum_{i\in I^{(2)}} E_i 
- \inv[\alpha]{2}\sum_{i\in I^{(4)}}(E_i^{(1)} + E_i^{(3)})
- \beta\sum_{i\in I^{(4)}}E_i^{(2)},
$$
where we can restrict to $\alpha,\beta\in\{0,\dots,3\}$. Then 
${\|B_4\|^2\over32}\in\Z$, which must hold by Lemma \ref{correctansatz},
iff $(\alpha,\beta)\in\{ (1,2),(1,3),(3,1),(3,2)\}$. 
For $(\alpha,\beta)=(3,1)$ there is no $E\in\Pi_4$ with 
$\langle B_4,E\rangle\equiv-1\mod4$.
We claim that the remaining three cases are equivalent by lattice automorphisms
in $O^+(H^2(X,\Z))$. Indeed, $(\alpha,\beta)=(1,3)$ turns into $(\alpha,\beta)
=(1,2)$ by
$$
E_i^{(1;3)}\longmapsto -E_i^{(1;3)}-E_i^{(2)}, 
\quad\quad
E_i^{(2)}\longmapsto E_i^{(1)}+E_i^{(2)}+E_i^{(3)},
$$
and $(\alpha,\beta)=(1,2)$ turns into $(\alpha,\beta)=(3,2)$ by
$$
E_i^{(1;3)}\longmapsto -E_i^{(1;3)}, \quad\quad
E_i^{(2)}\longmapsto E_i^{(1)}+E_i^{(2)}+E_i^{(3)}.
$$
Both maps induce identity on $\Pi_4^\ast/\Pi_4$ and hence can be 
trivially continued to elements of $O^+(H^2(X,\Z))$ by
\cite[Prop.\ 1.1]{ni80b}.
\epr
Lemmata  \ref{correctansatz}, \ref{bchoice},  and Theorem
\ref{grass} determine the desired embedding
$\wh\pi_\ast(H^{even}(T,\Z)^{G})\hookrightarrow 
H^{even}(X,\Z)$. Note that apart from the observation that
$\inv{2}\sum_{i\in\F_2^4} E_i\in\Pi_2$,
$\inv{2}\sum_{i\in I^{(4)}} (E_i^{(1)}+E_i^{(3)})\in\Pi_4$ etc.
the result for $B_{|G|}$ is independent of
the explicit calculations that led to Prop.\ \ref{kummerlattice}.
It is crucial to understand that for $E\in\Pi$ we found that in general
$E\not\in H^{even}(X,\Z)$. Lemma \ref{correctansatz} shows that
there is a lift $\wh E\in H^{even}(X,\Z)$ for every such vector. 
In particular, only $\wh E$  can have a geometrical meaning.
We lift
\begin{equation}\label{bz4}
\begin{array}{rcl}\ds
\wh B_{|G|}
:= B_{|G|} -\skp{B_{|G|}}{\wh\upsilon^0}\wh\upsilon
\>=\> B_{|G|} + \inv[\|B_{|G|}\|^2]{{|G|}} \wh\upsilon\e
\>=\> B_{|G|} - 2|G|\, \wh\upsilon
\quad\in\;\; H^{even}(X,\Z)
\end{array}
\end{equation}
to find that for generators of $\wh\pi_\ast x_T$
\begin{equation}\label{bfield}\begin{array}{rcl} 
\displaystyle
\mbox{with }\quad
B\;:=\;\inv{\sqrt{|G|}}B_T+\inv{{|G|}}\wh B_{|G|}: \ed
\fa\sigma\in\Sigma_T:\quad\quad
\sigma-\langle\sigma,B_T\rangle\upsilon
\>=\> \sigma-\langle \sigma,B\rangle\wh{\upsilon},\ed
\inv{\sqrt{|G|}}\left(\upsilon^0+B_T+\left(V_T-\inv[\|B_T\|^2]{2}\right)
\upsilon\right)
\>=\> \wh{\upsilon}^0+B
+\left(\inv[V_T]{{|G|}}-\inv[\left\|B\right\|^2]{2}
\right)\wh{\upsilon}.
\end{array}\end{equation}
Compare with Theorem \ref{grass} to see that this proves
\begin{envis}{Theorem}{z4emb}
Let $(\Sigma_T,V_T,B_T)$ denote a geometric interpretation of a toroidal 
nonlinear sigma model on the torus $T$ that admits a $G$ symmetry,
$G\subset SU(2)$ not containing non-trivial
translations; all such $G$ actions are specified
in \req{notations}. Then its image
$x\in{\cal T}^{4,20}$ under the
$G$ orbifold procedure has geometric interpretation
$(\Sigma,V,B)$ where $\Sigma\in{\cal T}^{3,19}$ is found as described by
Prop.\ \ref{kummerlattice},  $V={V_T\over {|G|}}$,
and $B={1\over\sqrt{|G|}}B_T+{1\over {|G|}}\wh B_{|G|}$,
$\wh B_{|G|}\in H^{even}(X,\Z)$ as in \req{bz4}
with $\skp{B_{|G|}}{E_s^{(j)}}=\skp{\wh B_{|G|}}{\wh E_s^{(j)}}$ 
the $|G\colon\!\!G^\prime|$--fold 
coefficient of $E_s^{(j)}$ in the highest root
of $\Gamma_s$,   $s\in\m S$   a $G^\prime\subset G$ type fixed point 
(see Lemma \ref{bchoice}).
\end{envis}\indent
With \req{bfield} we can confirm that the
primitiveness of $\sqrt{|G|}\upsilon,
\sqrt{|G|}\upsilon^0$ in $H^{even}(X,\Z)$ indeed follows from our requirement
that orbifold conformal field theories are consistently embedded in
$\m M^{K3}$. This can either be seen from consistency with the symmetry
$\upsilon\leftrightarrow\upsilon^0$ on $\m M^{tori}$ or the rescaling
of $B_T$ under the embedding that follows from \req{bfield}. Namely, assuming
$\inv{\sqrt{|G|}}\upsilon^0\in H^{even}(X,\Z)$ and hence
$B_{|G|}=\wh B_{|G|}=0$ in \req{newmlattice},
\req{bfield} directly leads to a contradiction:
$B_T$ and $B_T+\lambda$, $\lambda\in H^2(T,\Z)$, give equivalent 
torus theories by \req{modsp}, but $B=B_T/\sqrt{|G|}$, and 
$B=(B_T+\lambda)/\sqrt{|G|}$ in general do not give equivalent $K3$ theories. 
Not so if we use \req{bfield} with
$\wh B_{|G|}$ as determined
above, by the results of Prop.\ \ref{kummerlattice}, since then
the shift by $\lambda$ can be compensated by a lattice automorphism. 
Any other
ansatz with non-primitive $\sqrt{|G|}\upsilon,
\sqrt{|G|}\upsilon^0$ leads to an analogous contradiction by the methods
presented in Lemma \ref{bchoice}.
\section{Discussion}\label{conc}
Let us summarize the results of this work: 
By Prop.\ \ref{kummerlattice} and
Theorem \ref{z4emb}, for all orbifold constructions
of $K3$ obtained from non-translationary groups, 
the precise location of the corresponding 
orbifold conformal field theories  within the moduli space
$\m M^{K3}$ of theories associated to $K3$
has been determined. 
To arrive at Prop.\ \ref{kummerlattice}, we have presented a 
technique to calculate the generalization of the Kummer lattice to
all these orbifolds. The fact that the  components 
of the exceptional divisors of the blow up do not generate primitive
sublattices of $H_2(X,\Z)$ distinguishes our compact $X$ from
the minimal resolution of ${\C^2/G}$. 
The explicit results listed in Prop.\ \ref{kummerlattice} 
should allow a detailed analysis of D--branes on 
orbifold limits of $K3$
in the spirit of \cite{rawa98,ber99}.
Sect.\ \ref{quantum} contains an elementary new proof for the fact that 
the orbifold procedure forces fixed values on the B--field of the orbifold
conformal field theory in direction of the exceptional divisors  
\cite{as95}. Our proof is mostly independent of the technical 
discussion in Sect.\ \ref{geometry}. It merely uses the known description
of the moduli space $\m M^{K3}$ \cite{asmo94} and shows that the B--field
flux can be interpreted as artifact from a consistent embedding of 
orbifold conformal field theories in $\m M^{K3}$. We also prove that our
consistency requirement already fixes the B--field values uniquely 
up to lattice automorphisms, and we
are able to read them off explicitly. For the cyclic groups, we are in 
agreement with \cite{as95} ($G=\Z_2$), and with
\cite{do97,blin97} where mass formulae and tadpole cancellation conditions for
D--branes were used; the author did not find the explicit results for 
the binary dihedral and tetrahedral
groups in the literature.

Theorem \ref{z4emb} indicates a connection to the classical
McKay correspondence \cite{mk80,mk81}, which has also inspired very  
recent work in
the physics literature \cite{dido00,goja00,to00,ma00}. 
All of the latter publications  concentrate on higher
dimensional cases where large volume limits are used, though. Consider the 
local picture near any of our fixed points $s\in\m S$. Without loss of
generality $s=0$, and we are studying the minimal resolution $Y$ of
$\C^2/G^\prime, G^\prime\subset G$. 
By Theorem \ref{z4emb}, the contribution from this fixed point
to $\wh B_{|G|}$ is 
$|G\colon\!\!G^\prime|\,\wh B_{|G|}^s$ with
$\wh B_{|G|}^s:=\sum_j n_s^{(j)} (\wh E_s^{(j)})^\ast$, where 
$\sum_j n_s^{(j)} \wh E_s^{(j)}$ is the highest root of $\wh \Gamma_s$, and 
$\{ (\wh E_s^{(j)})^\ast\}\subset\wh\Gamma_s\otimes\Q$ 
denotes the dual basis of the fundamental system
$\{ \wh E_s^{(j)}\}$ of $\wh \Gamma_s$. Recall that the $j^{th}$ node 
in the extended
Dynkin diagram of $G^\prime$ labels an irreducible representation $\rho_j$ of
$G^\prime$ of dimension $n_s^{(j)}$, where $n_s^{(0)}=1$, and $\rho_0$ is the
trivial representation. 
On the other hand, the McKay correspondence
as proven in \cite{gsve83,kn85,arve85} states that $(\wh E_s^{(j)})^\ast$ is
the first Chern class of a locally free sheaf on $Y$ that is built from
the associated bundle on $\C^2/G^\prime-\{0\}$ 
given by $\rho_j$. Since the regular
representation $\rho$ of $G^\prime$
decomposes as $\rho=\sum_j n_s^{(j)} \rho_j$,
$\wh B_{|G|}^s$ is the first
Chern class of the extension  to $Y$ of 
$\pi_\ast \m O_{\C^2-\{0\}}(\C^{|G^\prime\!|})$ with regular
$G^\prime$ action on $\C^{|G^\prime\!|}$. It appears reasonable to assume that 
similarly to \cite{ko90}
the construction of \cite{gsve83,kn85}
can be carried over to $X$ by gluing appropriate sheaves near each fixed point
in a deformation of $X$ and taking the orbifold limit. For the present case
this in fact follows
from the results in \cite{bkr99}.
Then $\wh B_{|G|}$ is the first Chern class of a sheaf
$\m E\rightarrow X$ obtained from 
$\pi_\ast\m O_{T-\wt{\m S}}(\C^{|G|})$ by continuation. 
For $G=\Z_M, M\in\{2,3,4,6\}$, we find that the corresponding Mukai
vector \cite{mu84,mu87} obeys
\begin{eqnarray}\label{mukai}
\displaystyle
ch(\m E) \sqrt{\wh A(X)}
\>=\> [\rk\m E]\, \wh\upsilon^0 +c_1(\m E) 
+ \left[ (c_2-\inv{2}c_1^2)(\m E)[X] + \inv[\rk\m E]{48} p_1(X) \right]
\wh\upsilon\nonumber\e
\>=\> |G|\,\wh\upsilon^0 + \wh B_{|G|} + |G|\,\wh\upsilon
\,\stackrel{\req{newmlattice},\req{bz4}}{=}\, \sqrt{|G|}\upsilon^0
\,=\,\wh\pi_\ast \upsilon^0\e
\>\>\Longrightarrow \quad\quad 
\wh\pi^\ast \left( ch(\m E) \sqrt{\wh A(X)}\right)
= |G| \upsilon^0.\nonumber
\end{eqnarray}
Since it only remains to be shown that $(c_2-\inv{2}c_1^2)(\m E)[X]=2|G|$
for binary dihedral and tetrahedral
$G$, too, we conjecture \req{mukai} to hold in general.
In $H^{even}(T,\Z)$, 
$|G|\upsilon^0$ is the Mukai vector of a flat bundle of rank $|G|$
that naturally carries the regular representation $\rho$ of $G$ on the
fibers, yielding a $G$ equivariant flat bundle. Hence \req{mukai} 
is in exact agreement
with the McKay correspondence. We regard this as confirmation of  
Theorem \ref{z4emb}, though
Mukai vectors do not capture any information on $G$ equivariance, the 
basic ingredient of the McKay correspondence. 
That we need to choose the regular representation on the fiber has to 
do with our choices on the representative of $\wh B_{|G|}$, or more
precisely $\wh\upsilon^0$, above. Namely, at the end of Sect.\ \ref{quantum}
we have remarked that a shift of the B-field on the underlying torus theory 
by an integral form induces an integral shift  of $\wh\upsilon^0$
by some $b\in\wh\Pi_N/\oplus_s\wh\Gamma_s$, that 
changes the representation on our bundle of rank $|G|$ above.
This freedom of choice is readily checked to correspond to the
freedom of coordinate choice on the space of $G$ equivariant flat
bundles on the underlying torus $T$.

It is tempting to search for a direct
geometric interpretation of our methods:
Recall the geometric picture of Sect.\ \ref{geometry}. Here, the key
point was the construction of homology classes of type III.
which  arise from non-generic torus cycles by removing the branch locus
and then taking a single sheet of an \'etale covering. Similarly, the fact
that $\sqrt{|G|}\upsilon^0-B_{|G|}$ in \req{newmlattice} splits into
$|G|$ lattice vectors should be interpreted such that
$B_{|G|}$ corresponds to   the branch
locus of a $|G|\colon\!\!1$ branched covering of $X$. 
For $G=\Z_2$ such a covering exists \cite[(9)]{ni75}, and
for the cyclic groups similar ones have been 
constructed\footnote{I thank Claus Hertling 
for his explanations
on this point and for prodding me to the relevant literature.}
\cite{ti99}, but in general not of type $|G|\colon\!\!1$. Moreover,
we are lacking a precise mathematical formulation
for the mixing of degrees in $H^{even}(X,\Z)$ that would be needed for
such an interpretation. The determination of the exact form of $B_{|G|}$
from this approach also remains under investigation, 
though intuitively the characterization
of $B_{|G|}$ directly relates to 
that for the $E_s$ in III., Sect.\ \ref{geometry}.

However, since our lattice calculations imply an interpretation  
in this spirit, it shall be interesting to find the appropriate mathematical
framework.
\begin{noteadd}
After completion of this work we learned that the Kummer type lattices for
cyclic groups have already been determined by J.\ Bertin
in \cite{be88}. 
Prop.\ \ref{kummerlattice} is in agreement with these results on 
Abelian orbifold
constructions of $K3$.
\end{noteadd}
\appendix
\section{Lattices}\label{lattices}
The following material is taken from 
\cite{ni80b,ni80,mo84}.

A lattice $\Gamma\subset\R^{p,q}$ is called \textsl{integral}, 
if the associated
symmetric bilinear form is an integral form. It is \textsl{even}, if the 
associated quadratic form is
even.  By $\Gamma(N)$ we denote the same
$\Z$ module as $\Gamma$, but with quadratic form rescaled by a factor of $N$.
The \textit{discriminant} $\disc(\Gamma)$ is the determinant of the associated bilinear form on $\Gamma$.  The lattice $\Gamma$ is 
\textit{nondegenerate} if $\disc(\Gamma)\neq0$, and \textit{unimodular} if 
$|\!\disc(\Gamma)|=1$. If $\Gamma$ is a nondegenerate integral lattice, then 
$\disc(\Gamma) = |\Gamma^\ast\colon \Gamma|$,
where $\Gamma^\ast$ denotes the dual lattice of $\Gamma$
and $\Gamma\hookrightarrow\Gamma^*$ by using the bilinear form on $\Gamma$.
The \textit{signature} $(p,q)$ of $\Gamma$ is the 
multiplicity of the eigenvalues $(+1,-1)$ for the induced quadratic form on 
$\Gamma\otimes\R$. The \textit{discriminant form} $q_\Gamma$
associated to an even lattice
$\Gamma$ is the map $q_\Gamma: \Gamma^\ast/\Gamma\rightarrow
\Q/2\Z$ which is induced by the quadratic form of $\Gamma$,
together with the induced symmetric bilinear form on $\Gamma^\ast/\Gamma$
with values in $\Q/\Z$. 

The examples of even unimodular lattices most frequently used in our work
are the \textit{hyperbolic lattice} $U$ with quadratic form given by
$$\left( \begin{array}{cc} 0&1\\1&0 \end{array}\right),$$ and the lattice
$E_8$ with quadratic form given by the Cartan matrix of $E_8$. Moreover,
one has
\begin{envis}[\cite{mi58}]{Theorem}{onelatt}
If $\Gamma$ is an even unimodular lattice with signature $(p,p+\delta)$, 
$p>0$, $\delta\geq0$, then 
$$
\delta\equiv 0(8), \quad \Gamma\cong 
\Gamma^{p,p+\delta}:=U^p\oplus( E_8(-1) )^{\delta/8}.
$$
\vspace{-0.75em}\end{envis}\indent
A sublattice $\Lambda\subset\Gamma$ is \textit{primitive} iff
$\Gamma/\Lambda$ is free. A vector $\lambda\in\Gamma$ is primitive,
if $\Lambda:=\Z\lambda\subset\Gamma$ is primitive.

Embeddings of primitive sublattices in unimodular
lattices are characterized by
\begin{envis}[\mbox{\cite[Prop.\ 1.6.1]{ni80b}, 
\cite[\para1]{ni80}}]{Theorem}{lattemb}
Let $\Lambda$ denote  a primitive nondegenerate sublattice of an even
unimodular lattice $\Gamma$. Then the embedding $\Lambda\hookrightarrow\Gamma$
with $\Lambda^\perp\cap\Gamma\cong \m V$ is specified by an isomorphism 
$\gamma:\Lambda^\ast/\Lambda\rightarrow {\m V}^\ast/{\m V}$, such
that for the discriminant forms  $q_\Lambda=-q_{\m V}\circ\gamma$. Moreover,
$$
\Gamma\cong\left\{ (\lambda,v)\in \Lambda^\ast\oplus {\m V}^\ast\mid
\gamma(\qu\lambda)=\qu v \right\},
$$
where $\qu l$ denotes the projection of $l\in L^\ast$ onto $L^\ast/L$.
\end{envis}\indent\vspace*{-0.5em}
\section{Grassmannians}\label{grassmann}
By $\cal T^{a,b}$ we  denote the \textit{Grassmannian of oriented
positive 
definite subspaces $W\subset\R^{a,b}$ with} $\dim W=a$. Hence
$$
\cal T^{a,b} \cong O^+(a,b)/SO(a)\times O(b),
$$
where $O(a,b)=O(\R^{a,b})$ and analogously for $O^+(a,b)$, $SO(a,b)$,
$O(a,b)$, and $O(a)=O(a,0)$ etc.
Here, for any vector space $W$ with scalar product, 
$O(W)$ denotes the group of orthogonal transformations of $W$. Its subgroup
$O^+(W)$ contains all elements that do not interchange the two components
of the space of maximal positive definite subspaces of $W$. Note that
for positive definite $W$, $SO(W)=O^+(W)$.
For a lattice $\Gamma\subset W$ the group
$O(\Gamma)$ is the group of lattice automorphisms of $\Gamma$, and
$O^+(\Gamma)=O(\Gamma)\cap O^+(W)$ etc.

With the techniques of \cite{bose73} one shows:
\begin{envis}{Theorem}{grass}
For  $a,b\in\N$, there  is an isomorphism
$$
\cal T^{a+1,b+1}\cong \cal T^{a,b}\times\R^+\times\R^{a,b},
$$
which is specified by the choice of two null vectors 
$\upsilon,\upsilon^0\in\R^{a+1,b+1}$ with $\skp{\upsilon}{\upsilon^0}=1$
such that $\R^{a,b}\perp\upsilon,\upsilon^0$, and $\m T^{a,b}$ is built on
$\R^{a,b}$ in the above product.
Explicitly, we have
\begin{eqnarray}\label{asmois}
x\longmapsto (\Sigma,V,B) \quad\Longleftrightarrow\quad
x&=& \spann_\R\left(\,\xi(\Sigma),\, \upsilon^0+B+(V-{B^2/2})\upsilon\,\right),
\nonumber\\
\xi(\sigma)&=&\sigma-\skp{B}{\sigma}\upsilon.
\end{eqnarray}
The above isomorphism induces the structure of a warped product on 
$\cal T^{a,b}\times\R^+\times\R^{a,b}$.
\end{envis}\indent\vspace*{-0.5em}
%
\newcommand{\etalchar}[1]{$^{#1}$}
\def\polhk#1{\setbox0=\hbox{#1}{\ooalign{\hidewidth
  \lower1.5ex\hbox{`}\hidewidth\crcr\unhbox0}}}
\providecommand{\bysame}{\leavevmode\hbox to3em{\hrulefill}\thinspace}


\begin{thebibliography}{ABD{\etalchar{+}}76}

\bibitem[ABD{\etalchar{+}}76]{aetal76}
{\sc M.~Ademollo, L.~Brink, A.~D'Adda, R.~D'Auria, E.~Na\-po\-li\-tano, S.~Sciuto,
  E.~Del Giudice, P.~Di Vecchia, S.~Ferrara, F.~Gliozzi, R.~Musto, and
  R.~Pettorino}, \emph{Supersymmetric strings and color confinement}, Phys.
  Lett. \textbf{B62} (1976), 105--110.

\bibitem[Art66]{ar66}
{\sc M.~Artin}, \emph{On isolated rational singularities of surfaces}, Amer. J.
  Math. \textbf{88} (1966), 129--136.

\bibitem[AV85]{arve85}
{\sc M.~Artin and J.-L. Verdier}, \emph{Reflexive modules over rational double
  points}, Math. Ann. \textbf{270} (1985), 79--82.

\bibitem[AM]{asmo}
{\sc P.S. Aspinwall and D.R. Morrison}, \emph{Mirror symmetry and the moduli
  space of $K3$ surfaces}, to appear.

\bibitem[AM94]{asmo94}
\leavevmode\vrule height 2pt depth -1.6pt width 23pt, \emph{String theory on K3
  surfaces}, in: Mirror symmetry, B.~Greene and S.T. Yau, eds., vol.~II, 1994,
  pp.~703--716; {\tt hep-th/9404151}.

\bibitem[Asp95]{as95}
{\sc P.S. Aspinwall}, \emph{Enhanced gauge symmetries and $K3$ surfaces}, Phys.
  Lett. \textbf{B357} (1995), 329--334; {\tt hep-th/9507012}.

\bibitem[Asp97]{as96}
\leavevmode\vrule height 2pt depth -1.6pt width 23pt, \emph{$K3$ surfaces and
  string duality}, in: Fields, strings and duality (Boulder, CO, 1996), World
  Sci. Publishing, River Edge, NJ, 1997, pp.~421--540; {\tt hep-th/9611137}.

\bibitem[Ber88]{be88}
{\sc J.~Bertin}, \emph{R\'eseaux de {K}ummer et surfaces $K3$}, Invent. Math.
  \textbf{93} no.~2 (1988), 267--284.

\bibitem[BI97]{blin97}
{\sc J.D. Blum and K.~Intriligator}, \emph{Consistency conditions for branes at
  orbifold singularities}, Nucl. Phys. \textbf{B506} (1997), 223--235; {\tt
  hep-th/9705030}.

\bibitem[BS73]{bose73}
{\sc A.~Borel and J.-P. Serre}, \emph{Corners and arithmetic groups}, Comment.
  Math. Helv. \textbf{48} (1973), 436--491, Avec un appendice: Arrondissement
  des vari\'et\'es \`a coins, par A. Douady et L. H\'erault.

\bibitem[BKR99]{bkr99}
{\sc T.~Bridgeland, A.~King, and M.~Reid}, \emph{Mukai implies McKay: the McKay
  correspondence as an equivalence of derived categories}, 
J. Amer. Math. Soc. \textbf{14} no.~3 (2001), 535--554; {\tt
  math.AG/9908027}.

\bibitem[BER99]{ber99}
{\sc I.~Brunner, R.~Entin, and Ch. R{\"o}melsberger}, \emph{D-branes on
  $T^4/\mathbb{Z}_2$ and T-Duality}, JHEP \textbf{9906:016} (1999); {\tt
  hep-th/9905078}.

\bibitem[Cec91]{ce91}
{\sc S.~Cecotti}, \emph{$N=2$ 
Landau-Ginzburg vs. Calabi-Yau $\sigma$-models:
  Non per\-tur\-ba\-tive aspects}, Int.\ J.\ Mod.\ Phys.\ \textbf{A6} (1991),
  1749--1813.

\bibitem[dBDH{\etalchar{+}}]{bdhkmms01}
{\sc J.\ de Boer, R.\ Dijkgraaf, K.\ Hori, A.\ Keu\-ren\-tjes, J.\ Morgan, 
D.R.\ Morrison, and S.\ Sethi}, \emph{Triples, flu\-xes, and strings},
Adv. Theor. Math. Phys. \textbf{4} no.~5 (2000); {\tt
  hep-th/0103170}.

\bibitem[DD]{dido00}
{\sc D.-E. Diaconescu and M.R. Douglas}, \emph{D-branes on stringy Calabi-Yau
  manifolds}; {\tt hep-th/0006224}.

\bibitem[Dij99]{di99}
{\sc R.~Dijkgraaf}, \emph{In\-stan\-ton strings and hy\-per\-kaeh\-ler
  geo\-me\-try}, Nucl. Phys. \textbf{B543} (1999), 545--571; {\tt
  hep-th/9810210}.

\bibitem[Dou97]{do97}
{\sc M.R. Douglas}, \emph{Enhanced gauge symmetry in M(atrix) theory}, JHEP
  \textbf{9707:004} (1997); {\tt hep-th/9612126}.

\bibitem[EOTY89]{eoty89}
{\sc T.~Eguchi, H.~Ooguri, A.~Taormina, and S.-K. Yang}, \emph{Superconformal
  algebras and atring compactification on manifolds with SU(n) holonomy}, Nucl.
  Phys. \textbf{B315} (1989), 193--221.

\bibitem[Fuj88]{fu88}
{\sc A.~Fujiki}, \emph{Finite automorphism groups of complex tori of dimension
  two}, Publ. Res. Inst. Math. Sci. \textbf{24} no.~1 (1988), 1--97.

\bibitem[GSV83]{gsve83}
{\sc G.~Gonzalez-Sprinberg and J.-L. Verdier}, \emph{Construction
  g\'eom\'etrique de la correspondance de {M}c{K}ay}, Ann. Sci. \'Ecole Norm.
  Sup. \textbf{16} no.~3 (1983), 409--449.

\bibitem[GJ01]{goja00}
{\sc S.~Govindarajan and T.~Jayaraman}, \emph{D-branes, Exceptional Sheaves and
  Quivers on Calabi-Yau manifolds: From Mukai to McKay}, Nucl. Phys.
  \textbf{B600} (2001), 457--486; {\tt hep-th/0010196}.

\bibitem[Ino76]{in76}
{\sc H.~Inose}, \emph{On certain Kummer surfaces which can be realized as
  non-singular quartic surfaces in $\mathbb P^3$}, J.~Fac.~Sci.~Univ.~Tokyo
  \textbf{Sec. IA 23} (1976), 545--560.

\bibitem[Kn{\"o}85]{kn85}
{\sc H.~Kn{\"o}rrer}, \emph{Group representations and the resolution of
  rational double points}, in: Finite groups---coming of age (Montreal, Que.,
  1982), Amer. Math. Soc., Providence, R.I., 1985, pp.~175--222.

\bibitem[Kob90]{ko90}
{\sc R.~Kobayashi}, \emph{Moduli of {E}instein metrics on a ${K}3$ surface and
  degeneration of type I}, in: K\"ahler metric and moduli spaces, Academic
  Press, Boston, MA, 1990, pp.~257--311.

\bibitem[LP81]{lope81}
{\sc E.~Looijenga and C.~Peters}, \emph{Torelli theorems for K3-surfaces},
  Compos.~Math. \textbf{42} (1981), 145--186.

\bibitem[May01]{ma00}
{\sc P.~Mayr}, \emph{Phases of supersymmetric D-branes on K{\"a}hler manifolds
  and the McKay correspondence}, JHEP \textbf{0101:018} (2001); {\tt
  hep-th/0010223}.

\bibitem[McK80]{mk80}
{\sc J.~McKay}, \emph{Graphs, singularities, and finite groups}, in: The Santa
  Cruz Conference on Finite Groups (Univ. California, Santa Cruz, Calif.,
  1979), Amer. Math. Soc., Providence, R.I., 1980, pp.~183--186.

\bibitem[McK81]{mk81}
\leavevmode\vrule height 2pt depth -1.6pt width 23pt, \emph{Cartan matrices,
  finite groups of quaternions, and {K}leinian singularities}, Proc. Amer.
  Math. Soc. \textbf{81} no.~1 (1981), 153--154.

\bibitem[Mil58]{mi58}
{\sc J.~Milnor}, \emph{On simply connected 4-manifolds}, in: Symposium
  Internacional de Topologia Algebraica, La Universidad Nacional Aut\'onoma de
  M\'exico y la UNESCO, 1958, pp.~122--128.

\bibitem[Mor]{primo}
{\sc D.R. Morrison}, private communication.

\bibitem[Mor84]{mo84}
\leavevmode\vrule height 2pt depth -1.6pt width 23pt, \emph{On K3 surfaces with
  large Picard number}, Invent. Math. \textbf{75} (1984), 105--121.

\bibitem[Muk84]{mu84}
{\sc S.~Mukai}, \emph{On the symplectic structure of the moduli spaces of
  stable sheaves over abelian varieties and $K3$ surfaces}, Invent. Math.
  \textbf{77} (1984), 101--116.

\bibitem[Muk87]{mu87}
\leavevmode\vrule height 2pt depth -1.6pt width 23pt, \emph{On the moduli space
  of bundles on $K3$ surfaces I}, in: Vector bundles on algebraic varieties,
  Tata Inst. Fund. Res., Bombay, 1987, pp.~341--413.

\bibitem[NS95]{nasu95}
{\sc M.~Nagura and K.~Sugiyama}, \emph{Mirror symmetry of K3 and torus}, Int.
  J. Mod. Phys. \textbf{A10} (1995), 233--252; {\tt hep-th/9312159}.

\bibitem[NW01]{nawe00}
{\sc W.~Nahm and K.~Wendland}, \emph{A hiker's guide to $K3$ -- Aspects of
  $N=(4,4)$ superconformal field theory with central charge $c=6$}, Commun.
  Math. Phys. \textbf{216} (2001), 85--138; {\tt hep-th/9912067}.

\bibitem[Nar86]{na86}
{\sc K.S. Narain}, \emph{New heterotic string theories in uncompactified
  dimensions $< 10$}, Phys. Lett. \textbf{169B} (1986), 41--46.

\bibitem[Nik75]{ni75}
{\sc V.V. Nikulin}, \emph{On Kummer Surfaces}, Math. USSR Isv. \textbf{9}
  (1975), 261--275.

\bibitem[Nik80a]{ni80b}
\leavevmode\vrule height 2pt depth -1.6pt width 23pt, \emph{Finite automorphism
  groups of Kaehler K3 surfaces}, Trans. Mosc. Math. Soc. \textbf{38} (1980),
  71--135.

\bibitem[Nik80b]{ni80}
\leavevmode\vrule height 2pt depth -1.6pt width 23pt, \emph{Integral symmetric
  bilinear forms and some of their applications}, Math. USSR Isv. \textbf{14}
  (1980), 103--167.

\bibitem[P{\v{S}}{\v{S}}71]{pss71}
{\sc I.I. Pjatecki{\uu\i}-{\v{S}}apiro and I.~R. {\v{S}}afarevi{\v{c}}},
  \emph{Torelli's theorem for algebraic surfaces of type $K3$}, Izv. Akad. Nauk
  SSSR Ser. Mat. \textbf{35} (1971), 530--572.

\bibitem[RW98]{rawa98}
{\sc S.~Ramgoolam and D.~Waldram}, \emph{Zero branes on a compact orbifold},
  JHEP \textbf{9807:009} (1998); {\tt hep-th/9805191}.

\bibitem[Sei88]{se88}
{\sc N.~Seiberg}, \emph{Observations on the moduli space of superconformal
  field theories}, Nucl. Phys. \textbf{B303} (1988), 286--304.

\bibitem[SI77]{shin77}
{\sc T.~Shioda and H.~Inose}, \emph{On singular K3 surfaces}, in: Complex
  Analysis and Algebraic Geometry, W.L. Bailey and T.~Shioda, eds., Cambridge
  Univ. Press, 1977, pp.~119--136.

\bibitem[Tib99]{ti99}
{\sc M.~Tib{\uu{a}}r}, \emph{Monodromy of functions on isolated cyclic
  quotients}, Topology Appl. \textbf{97} no.~3 (1999), 231--251.

\bibitem[Tom01]{to00}
{\sc A.~Tomasiello}, \emph{D-branes on Calabi-Yau manifolds and helices}, JHEP
  \textbf{0102:008} (2001); {\tt hep-th/0010217}.

\bibitem[Val34]{du34}
{\sc P.~Du Val}, \emph{On isolated singularities which do not affect the
  condition of adjunction}, Proc. Cambridge Phil. Soc. \textbf{30} (1934),
  453--465.

\end{thebibliography}
\end{document}